\documentclass[usenatbib,papersize=a4]{mn2e}

\usepackage{amsmath}
\usepackage{natbib}
\usepackage{epsfig}
\usepackage{txfonts}
\usepackage{pict2e}
\usepackage{subcaption}
\usepackage[usenames,dvipsnames]{color}
\usepackage{hyperref}

\hypersetup{
  colorlinks   = true, 
  urlcolor     = NavyBlue, 
  linkcolor    = blue, 
  citecolor   = blue 
}

\newcommand{\eV}{{\rm eV}}

\newcommand{\Kel}{{\rm K}}

\newcommand{\cm}{{\rm cm}}

\newcommand{\Mpc}{{\rm Mpc}}

\newcommand{\expf}[1]{{{\rm e}^{#1}}}

\newcommand{\kD}{k_{\rm D}}

\newcommand{\id}{{\,\rm d}}

\newcommand{\dk}{\ensuremath{\mathrm{d}k}}
\newcommand{\dlnk}{\ensuremath{\mathrm{d}\ln k}}

\newcommand{\beq}{\begin{equation}}   %

\newcommand{\eeq}{\end{equation}}   %

\newcommand{\beqa}{\begin{eqnarray}}   %

\newcommand{\eeqa}{\end{eqnarray}}   %

\newcommand{\beal}{\begin{align}}
\newcommand{\enal}{\end{align}}

\newcommand{\bspl}{\begin{split}}

\newcommand{\espl}{\end{split}}

\newcommand{\bsub}{\begin{subequations}}

\newcommand{\esub}{\end{subequations}}

\newcommand{\bmulti}{\begin{multline}}   %

\newcommand{\beqm}{\begin{mathletters}}   %

\newcommand{\eeqm}{\end{mathletters}}   %

\newcommand{\Ne}{N_{\rm e}}

\newcommand{\Te}{T_{\rm e}}

\newcommand{\Tg}{T_{\gamma}}

\newcommand{\sigT}{\sigma_{\rm T}}

\newcommand{\vek} [1]{\mbox{\boldmath${#1}$\unboldmath}}

\newcommand{\pot}[2]{#1 \times 10^{#2}}

\newcommand{\logstr}{_loglog}

\usepackage{hyperref}
\usepackage{grffile}
\usepackage{graphics}
\usepackage{booktabs}

\title[Magnetic heating across recombination]
{Magnetic heating across the cosmological recombination era: Results from 3D MHD simulations}

\author[Trivedi et al.]{Pranjal Trivedi$^{1,2}$\thanks{E-mail: pranjal.trivedi@hs.uni-hamburg.de}, Johannes Reppin$^{1}$\thanks{E-mail: johannes.reppin@hs.uni-hamburg.de}, Jens~Chluba$^{3}$\thanks{E-mail: jens.chluba@manchester.ac.uk} and Robi Banerjee$^{1}$
\\
$^{1}$Hamburger Sternwarte, University of Hamburg, Gojenbergsweg 112, 21029, Hamburg, Germany
\\
$^{2}$Department of Physics, Sri Venkateswara College, University of Delhi, 110021 New Delhi, India
\\
$^{3}$Jodrell Bank Centre for Astrophysics, Alan Turing Building, University of Manchester, Manchester M13 9PL, UK
}

\begin{document}

\date{\vspace{-10mm}} 

\maketitle

\begin{abstract}
The origin of cosmic magnetic fields is an unsolved problem and magnetogenesis could have occurred in the early Universe. 
We study the evolution of such primordial magnetic fields across the cosmological recombination epoch via 3D magnetohydrodynamic numerical simulations. 
We compute the effective or net heating rate of baryons due to decaying magnetic fields and its dependence on the magnetic field strength and spectral index. 
In the drag-dominated regime ($z \gtrsim 1500$), prior to recombination, we find no real heating is produced. 
Our simulations allow us to smoothly trace a new transition regime ($600 \lesssim z \lesssim 1500$), where magnetic energy decays, at first, into the kinetic energy of baryons. A turbulent velocity field is built up until it saturates, as the net heating rate rises from a low value at recombination to its peak towards the end of the transition regime.
This is followed by a turbulent decay regime ($z \lesssim 600$) where magnetic energy dissipates via turbulent decay of both magnetic and velocity fields while net heating remains appreciable and declines slowly.
Both the peak of the net heating rate and the onset of turbulent decay are delayed significantly beyond recombination, by up to 0.5 Myr (until $z\simeq 600-700$), for scale-invariant magnetic fields. 
We provide analytic approximations and present numerical results for a range of field strengths and spectral indices, illustrating the redshift-dependence of dissipation and net heating rates. These can be used to study cosmic microwave background constraints on primordial magnetic fields.
\end{abstract}


\section{Introduction}
\label{sec:Intro}
%
Magnetic fields are observed ubiquitously and play an important role in the physics of galaxies out to large scales in the Universe \citep{Beck2015,Han2017}. 
Microgauss strength magnetic fields have also been detected in high redshift galaxies \citep[e.g.][]{Bernet2008,Kronberg2008,Mao2017} and across scales of several megaparsecs in clusters of galaxies \citep{Clarke2001,Bonafede2010,Feretti2012}. 
There is also indirect evidence from $\gamma$-ray observations suggesting a lower limit of $B \sim 10^{-16}$ G on intergalactic magnetic fields that can fill most of the cosmic volume \citep{Neronov2010,Tavecchio2011,Dolag2011,Taylor2011,Dermer2011,Fermi2018}.

These observations motivate a primordial origin, i.e., from the very early Universe, for the seed magnetic field responsible for these observed large-scale fields. 
Theoretical models have been proposed for the generation of such a primordial magnetic field (PMF) which generically involves breaking of conformal invariance of the electromagnetic action. Primordial magnetogenesis could have occurred at inflation \citep{Turner1988,Ratra1992,Gasperini1995,Martin2008,Subramanian2010,Caprini2014,Sharma2017} or at a cosmological phase transition \citep{Vachaspati1991,Sigl1997,Grasso2001,Copi2008} 
However, a complete theory of magnetogenesis and the PMF's subsequent evolution remains an important unresolved issue in cosmology \citep[reviewed in ][]{Kandus2011,DurrerNeronov2013,Subramanian2016}.

The temperature and polarization anisotropies of the cosmic microwave background (CMB) are a sensitive probe of PMFs since magnetic fields induce metric perturbations as well as fluid perturbations via the Lorentz force \citep{Subramanian2006,Finelli2008, Paoletti2009, Shaw2010PMF}. 
Upper limits have been placed on the comoving PMF at megaparsec scales 
of order or just below $B\simeq1$~nG derived from comparison to \textit{Planck} observations, based on the CMB power spectrum \citep{Chluba2015PMF,Planck2015PMF,Zucca2017} and CMB non-Gaussianity \citep{Trivedi2012,Trivedi2014,Shiraishi2014}. 
Similar PMF upper limits exist from a range of other probes of magnetic fields including Faraday rotation of radio sources \citep{Pshirkov2016} and changes to the cosmic ionization history \citep{Sethi2005,Kunze2014,Kunze2015,Chluba2015PMF,Planck2015PMF}.

However, a significant gap of several orders of magnitude between the upper limits and the inferred lower limits on large-scale primordial magnetic fields remains. 
Also, many of the currently applied approximations for magnetic heating and dissipation that are used to incorporate the effect on the cosmic ionization history, remain rather unsatisfactory \citep[e.g., see discussions in][]{Chluba2015PMF, Planck2015PMF}. 
Thus, an improved understanding of PMF signatures is needed to clarify their current amplitude, evolution and origin. 
Motivated by the outstanding issues, we investigate the amount of decay and dissipation of a PMF and the resultant heating produced across the cosmological recombination era \citep{Zeldovich68, Peebles68, Sunyaev2009}, one of the clearly anticipated phases in the evolution of PMFs.

PMFs are expected to evolve in a qualitatively distinct fashion across different cosmological epochs. 
Damping of PMFs can occur via neutrino- or photon-sourced fluid viscosity, via the development of magnetohydrodynamic (MHD) turbulent decay or processes such as ambipolar diffusion \citep{Jedamzik1998,Subramanian1998,Banerjee2004}.

Prior to cosmological recombination, there exists a photon drag-dominated viscous regime for the baryon-photon fluid. 
Close to the recombination epoch, once the radiation drag has started to drop abruptly, the PMF can source significant fluid motions
. This induces turbulence in the baryon fluid which ultimately dissipates some fraction of the energy contained in the PMF via a turbulent cascade. 
The energy dissipated from the PMF can heat electrons and baryons to alter the thermal history \citep{Sethi2005,Kunze2014,Kunze2015,Chluba2015PMF}. We also note that a phase of MHD turbulent decay of the PMF can also occur subsequent to earlier phase transitions (eg. electroweak \cite{Kahniashvili2010}). 

In this work, we report MHD simulations that describe the transfer and subsequent dissipation of magnetic energy across the epoch of recombination. 
Previous work has shown that turbulent decay will be important for PMF evolution after recombination \citep{Sethi2005,Chluba2015PMF}. 
Here, using full 3D simulations, we are able to quantify both the sourcing of the velocity field, the development of turbulence and the decay of the PMF. 
Our simulations allow us to smoothly follow the PMF and baryon velocity field across three distinct regimes: the photon drag-dominated regime, a new transition regime around recombination and the post-recombination regime dominated by turbulent decay. 
In particular, we can trace the relatively rapid evolution of baryon velocities, magnetic fields and the heating rate (of baryons and electrons) due to magnetic decay during the transition regime. 
This enables us to connect the evolution of magnetic fields in the pre- and post-recombination epochs. 
We can then quantify the PMF decay and heating rate of the baryon fluid as a function of PMF parameters: amplitude and spectral index. 
The timescale for turbulence to develop after recombination also depends on the PMF parameters and this affects the epoch of peak heating caused by the PMF. 
The simulations and fits presented in this work refine our understanding of the evolution of the PMF and offer a way to potentially place improved constraints on the nature of primordial magnetic fields.

Quantifying the heating effect of PMF dissipation via decaying turbulence can be related to two further effects in the CMB caused by fluid heating. 
Heated electrons result in delayed recombination which in turn modifies the CMB power spectra \citep{Sethi2005, Chluba2015PMF, Kunze2015}. 
Secondly, the additional energy in electrons up-scatters photons to higher energies, producing a distortion in the CMB spectrum \citep{Jedamzik2000, Chluba2011therm}. 
These effects are important when placing improved constraints on the PMF and its origin. 
In addition, the effects of PMF dissipation and heating are also relevant for assessing the role of the PMF in structure formation \citep{Wasserman1978,Kim1996,Subramanian1998,Sethi2005}, 21-cm signals \citep{Sethi2009,Schleicher2009,Shiraishi2014_21cm}, reionization signals \citep{Sethi2009,Bowman2018} and primordial chemistry \citep{Schleicher2008,Schleicher2008b}, thus warranting careful treatments of the process. 
Our simulation outputs can also be used as initial conditions for the small-scale velocity field in structure formation simulations addressing the effects of PMFs.

The subsequent sections are organized in the following manner: In Section~\ref{sec:av_B} we give a brief description of the magnetic field properties in the early Universe and present an overview of the their evolution. 
In Section~\ref{sec:analytics} we revisit the analytical treatment of the different evolutionary stages, specifically for the free-streaming damping epoch and the turbulent decay after recombination. 
The MHD simulations are described in Section~\ref{sec:simulations}, where  
we explain the numerical setup used to integrate the governing MHD equations and the techniques used for modeling the epoch of recombination. 
In Section~\ref{sec:Results} we present the results of the numerical simulations for magnetic field and velocity amplitudes, magnetic and kinetic power spectra and dissipation and net heating rates. 
Results are first described for the fiducial case with an initially near scale-invariant magnetic power spectrum. 
Subsequently, we present and analyze the trends in these physical results when varying the PMF parameters of the field amplitude and spectral index. 
We conclude the results section by providing a semi-analytical representation of the dissipation rate sourced by the decay of PMF in Section~\ref{sec:semi-analytics}. 
A discussion of our results and caveats follows in Section~\ref{sec:discussion} and conclusions are presented in Section~\ref{sec:conclusion}

\vspace{-0mm}
\section{Primordial stochastic magnetic field}
\label{sec:av_B}
We start by defining how the root mean square (rms) value of the primordial stochastic magnetic field (PMF) is related to the initial power spectrum at any moment. 
We assume that the initial power spectrum is given by $P_B(k)= A k^{n_B}$, $n_B>-3$. 
Then the r.m.s.{\it comoving} magnetic field strength over some Gaussian filtering scale $\lambda_{\rm f}$ is related to
\begin{align}
\label{eq:def_B_av}
B^2_{\rm f}=a^4 \mathcal{B}^2_{\rm f}=\int \frac{\id ^3k}{(2\pi)^2} P_B(k) \,\expf{-\lambda^2_{\rm f} k^2}=\frac{A\,\Gamma\left(\frac{n_B+3}{2}\right)}{4\pi^2 \lambda_{\rm f}^{n_B+3}},
\end{align}
where $\mathcal{B}^2_{\rm f}\propto (1+z)^4$ denotes the mean square field strength in proper coordinates.
Equation~\eqref{eq:def_B_av} is simply a definition without any physical input at this point. For example, $\lambda_{\rm f}=1\,\Mpc$ is frequently used to define $B_\lambda$ at large cosmological scales \citep[e.g.,][]{Planck2015PMF}.
One simple way to capture the evolution and damping of the rms magnetic field as a function of different epochs (e.g., for free-streaming damping) can be found by setting the filtering scale to $\lambda_{\rm f}=\sqrt{2}/k_{\rm d}(z)$, where $k_{\rm d}$ is a specified damping scale that depends on time. 
This is equivalent to saying that the initial power spectrum amplitude in Fourier space is modulated by a factor $\simeq \expf{-2k^2/k_{\rm d}^2(z)}$. We thus have
\begin{align}
\label{eq:barB}
\bar{B}^2=\frac{A\,\Gamma\left(\frac{n_B+3}{2}\right)}{4\pi^2}\frac{k_{\rm d}^{n_B+3} (z)}{2^{(n_B+3)/2}}.
\end{align}
In this way, the rms magnetic field strength is determined by the evolution of $k_{\rm d}(z)$. 
For near scale-invariant fields (e.g., $n_B=-2.9$), most of the energy is stored at the largest scales and even a significant evolution of $k_{\rm d}$ does not affect $\bar{B}^2$ very much. 
In contrast, for very blue initial PMF power spectra, most of the energy density is present at small scales and rapid loss is expected even if $k_{\rm d}(z)$ only evolves slowly. 
This simplified picture also provides a reasonable description of the magnetic field evolution in the turbulent decay phase, even if no exponential cut-off is expected at small scales, $k>k_{\rm d}(z)$ \citep{Jedamzik1998, Banerjee2004}.

We will discuss below the evolution of the rms magnetic field strength across the recombination era in more detail. 
It is expected that in the phase well after recombination but before reionization ($10\lesssim z \lesssim 10^3$) the magnetic field energy density changes relatively slowly \citep{Banerjee2004}. 
In this regime, {\it turbulent decay} of magnetic fields leads to a logarithmic evolution of the magnetic field strength with redshift. 
In particular for near scale-invariant initial power spectra, this implies a close to constant comoving magnetic field energy density, $\rho_{B}= a^{-4} \bar{B}^2/8\pi$. 

For convenience, in our analytic treatment we will normalize all magnetic field power spectra with respect to the redshift at which turbulent decay is expected to begin in the post-recombination era, $z_{\rm td}\simeq 10^3$. 
We will determine $z_{\rm td}$ more precisely below, but with this definition we approximately have
\begin{align}
\label{eq:barB_norm}
\bar{B}^2\approx B_0^2
\left(\frac{k_{\rm d} (z)}{k_{\rm td}}\right)^{n_B+3},
\end{align}
with $B_0\equiv\bar{B}(z_{\rm td})$ and $k_{\rm td}=k_{\rm d} (z_{\rm td})$. 
We now discuss the pre-recombination heating, which is dominated by free-streaming damping at $z\gtrsim z_{\rm td}$, and the post-recombination heating separately, improving upon the previous treatments in the literature.

\section{Analytical considerations in the different regimes}
\label{sec:analytics}
The damping of magnetic fields and heating of the medium occurs in alternating phases of turbulent and viscous regimes \citep{Jedamzik1998, Banerjee2004}. 
In this section, we briefly recap and attempt to improve previous analytic approximations in the different eras. 
We start with the phase just before recombination and then discuss the turbulent decay regime in the post-recombination epoch ($z\lesssim 10^3$). 
These approximations are then compared with the numerical results in Sect.~\ref{sec:Results}.

\vspace{-2mm}
\subsection{Free streaming damping}
\label{sec:free-damp}
At sufficiently small scales (below the photon mean free path) and also at later redshifts after recombination, photons stream freely relative to the baryons. 
In this case, the baryon Alfv\'en speed, $V_{\rm bA}$ (in units of the light speed), in a magnetized medium is given by 
\begin{align}
\label{def:Alfven_speed_etc}
V_{\rm bA}&=\frac{a^{-2}\bar{B}}{\sqrt{4\pi \rho_{\rm b}}}\equiv \sqrt{\frac{3}{2 R}}\,\sqrt{\frac{\rho_B}{\rho_\gamma}}
\approx\frac{\pot{3.8}{-4}}{\sqrt{R}}\left[\frac{\bar{B}}{1\,{\rm nG}}\right]
\nonumber\\
&
\approx \pot{1.5}{-5} (1+z)^{1/2}\left[\frac{\Omega_{\rm b} h^2}{0.022}\right]^{-1/2}\left[\frac{T_0}{2.726\,\Kel}\right]^{2}
\left[\frac{\bar{B}}{1\,{\rm nG}}\right]
\nonumber\\
R&=\frac{3\rho_{\rm b}}{4\rho_\gamma}\approx \frac{666}{1+z}\,\left[\frac{\Omega_{\rm b} h^2}{0.022}\right]\left[\frac{T_0}{2.726\,\Kel}\right]^{-4}
\nonumber\\
\rho_B&=\frac{a^{-4} \bar{B}^2}{8\pi}\approx \pot{9.5}{-8} \left[\frac{\bar{B}}{1\,{\rm nG}}\right]^2\,\rho_\gamma
\nonumber\\
\rho_\gamma&\approx 0.26 \,\eV\,\cm^{-3} \left[\frac{T_0}{2.726\,\Kel}\right]^{4} (1+z)^4,
\end{align}
where $\bar{B}$ is the rms magnetic field strength, $R$ the baryon-loading of the fluid, $\rho_B$ and $\rho_\gamma$ the rms magnetic and photon energy densities respectively. 
The free-streaming photons exert a small drag force, $\vek{F}_{\rm D}\approx -\frac{4}{3} \Ne \sigT \rho_\gamma \vek{\varv}$, on the baryons, mediated by the occasional scatterings of photons with the free electrons. 
This leads to damping of magnetic fields at very small scales\footnote{We denote comoving scales with $\lambda$ and proper scales with $\ell=a \lambda$.}, $\lambda\lesssim \lambda_{\rm fs}$. 
These interactions source small-scale photon perturbations, which gradually damp due to free-streaming mixing \citep[e.g., see][for recent study of generation of small-scale fluctuations by PMFs in clusters]{Minoda2017}.

\subsection{Estimates for the free-streaming scale}
\label{sec:free-streaming-scale}
\begin{figure}
\centering
\includegraphics[width=0.92\columnwidth]{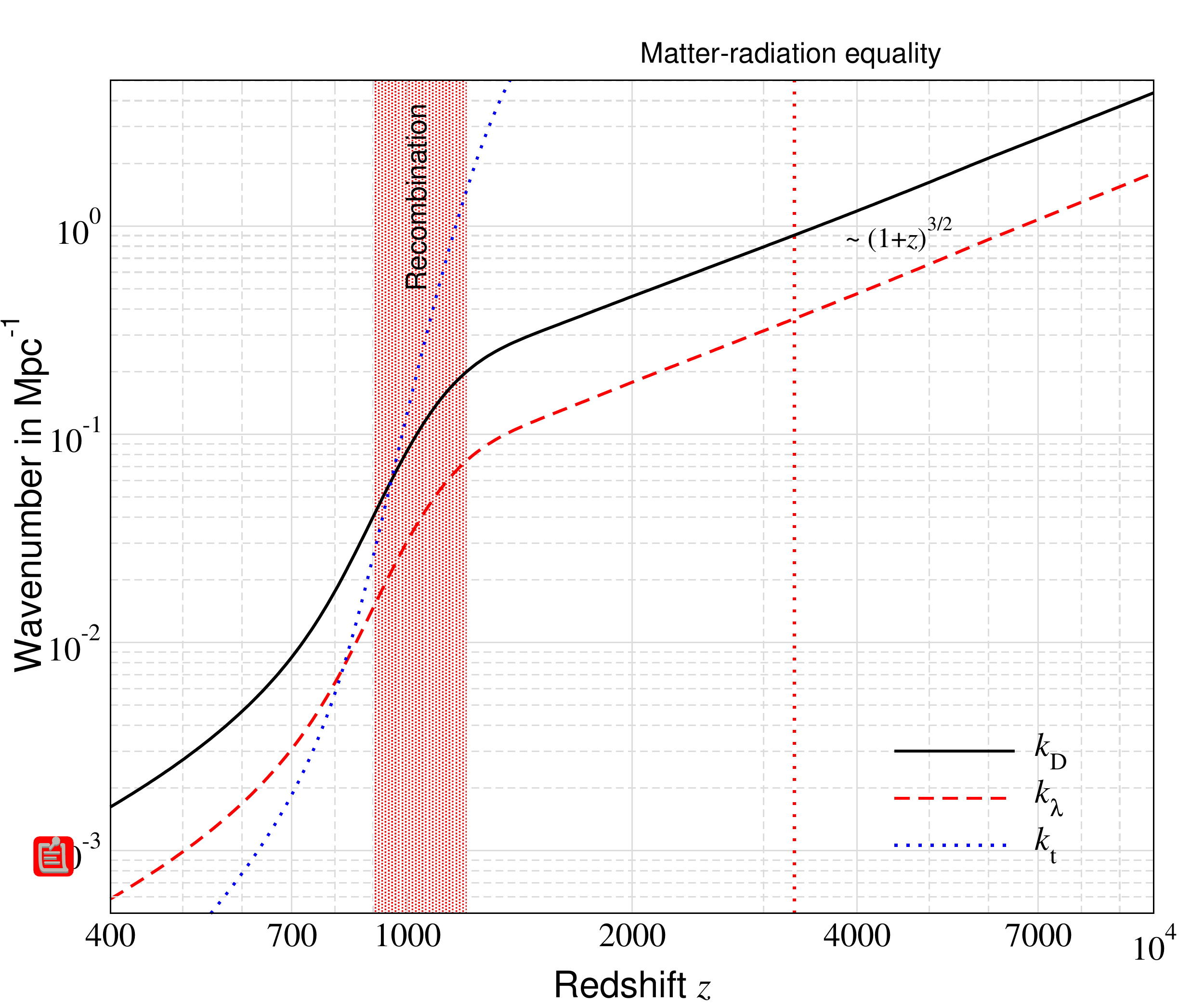}
\caption{Illustration for redshift dependence of some of the relevant scales. Here, $\kD$ and $k_\lambda$ are defined by Eq.~\eqref{eq:k_lambda_def} and \eqref{eq:k_D_gamma}, respectively. We also introduced $k_{\rm t}= 2\pi \Ne \sigT a/\sqrt{R}$ for convenience.}
\label{fig:k_values}
\end{figure}

The free-streaming scale is approximately determined by the condition \citep{Subramanian2016}
\begin{align}
\label{def:k_td}
\frac{1}{k_{\rm fs}^2}\approx \int \frac{V_{\rm bA}^2 R \id t}{\Ne \sigT a^2}\approx \pot{1.4}{-7} 
\left[\frac{B_0}{1\,{\rm nG}}\right]^2 \int \left(\frac{k_{\rm fs}}{k_{\rm td}}\right)^{n_B+3}\!\frac{\id t}{\Ne \sigT a^2},
\end{align}
where the baryon Alfv\'en speed, $V_{\rm bA}$, baryon loading factor of the fluid, $R$, and magnetic energy density, $\rho_B$, defined in Eq.~\eqref{def:Alfven_speed_etc} give $V_{\rm bA}^2 R = 3\rho_B/ 2\rho_\gamma$. For convenience we introduce the integral
\begin{align}
\label{eq:k_lambda_def}
\frac{1}{k^2_\lambda}=\int_0^t \frac{\id t'}{\Ne \sigT a^2}=\int^\infty_{z} \frac{\id z}{\Ne \sigT a H(z)},
\end{align}
which can be computed numerically and only depends on the standard background cosmology. It is approximately given by (see Fig.~\ref{fig:k_values}), $k_\lambda\approx \sqrt{8/45}\, \kD$, where the standard photon diffusion damping scale, $\kD$, is defined by \citep{Weinberg1971, Kaiser1983}
\begin{align}
\label{eq:k_D_gamma}
\frac{1}{k_{\rm D}^2}\approx \frac{8}{45}\int^\infty_z \frac{\id z}{\Ne \sigT a H}\left[\frac{1}{1+R}+\frac{15}{16}\frac{R^2}{(1+R)^2}\right],
\end{align}
for shear viscosity $\eta_\gamma=\frac{16}{45}\frac{\rho_\gamma}{ \Ne \sigT}$, which includes the effect of polarization\footnote{Without this effect one has $\eta=\frac{4}{15}\frac{\rho_\gamma}{ \Ne \sigT}$ \citep{Weinberg1971}.}. We have $k_{\rm D}\approx \pot{4.1}{-6}(1+z)^{3/2}\,\Mpc^{-1}$ in radiation-dominated era, such that $k_\lambda\approx \pot{1.8}{-6}(1+z)^{3/2}\,\Mpc^{-1}$. In the matter-dominated regime, one has $k_\lambda\propto (1+z)^{5/4}$; however, in our computation we use the full numerical result, which captures the rapid changes of $k_{\rm D}$ and $k_\lambda$ around $z\simeq 10^3$ (Fig.~\ref{fig:k_values}).

By separating variables in Eq.~\eqref{def:k_td} and solving for $k_{\rm fs}$ we obtain
\begin{align}
\label{def:kfs_rad_II}
k_{\rm fs}&\approx  k_{\rm td}\,\left( \pot{1.4}{-7}\,\frac{(n_B+5)}{2}
\left[\frac{B_0}{1\,{\rm nG}}\,\frac{k_{\rm td}}{k_\lambda}\right]^2\right)^{-\frac{1}{n_B+5}}
=k_{\rm td}\,\left[\frac{k_\lambda}{k_\lambda(z_{\rm td})}\right]^{\frac{2}{n_B+5}}
\nonumber\\
k_{\rm td}&=\sqrt{\frac{2}{n_B+5}}\,\frac{k_\lambda(z_{\rm td})}{\pot{3.8}{-4}}
\left[\frac{B_0}{1\,{\rm nG}}\right]^{-1},
\end{align}
where we used $k_{\rm fs}\approx k_{\rm td}$ at $z_{\rm td}$. Usually, the time-dependence of $k_{\rm fs}/k_{\rm td}$ is omitted, such that the factor $\left(k_{\rm fs}/k_{\rm td}\right)^{n_B+3}$ in Eq.~\eqref{def:k_td} can taken outside the integral. This yields a slightly different solution
\begin{align}
\label{def:kfs_rad_II_old}
k^*_{\rm fs}&\approx  k^*_{\rm td}\,\left( \pot{1.4}{-7}
\left[\frac{B_0}{1\,{\rm nG}}\,\frac{k^*_{\rm td}}{k_\lambda}\right]^2\right)^{-\frac{1}{n_B+5}}
=k^*_{\rm td}\,\left[\frac{k_\lambda}{k_\lambda(z_{\rm td})}\right]^{\frac{2}{n_B+5}}
\nonumber\\
k^*_{\rm td}&=\frac{k_\lambda(z_{\rm td})}{\pot{3.8}{-4}}
\left[\frac{B_0}{1\,{\rm nG}}\right]^{-1},
\end{align}
which does not include any dependence of $k_{\rm fs}$ and $k_{\rm td}$ on the PMF spectral index. To fix $k_{\rm fs}$ we still need to estimate $z_{\rm td}$. However, it is already clear that $z_{\rm td}\simeq 10^3$, thus, $k_\lambda(z_{\rm td})\simeq 0.01-0.1\,\Mpc^{-1}$, which puts us into the regime $k_{\rm td}\simeq 26-260\,\Mpc^{-1}$ for $n_{B}\simeq -3$ and $B_0\simeq 1\,{\rm nG}$. 
Due to resolution issues and to avoid shocks, in our simulations we are only able to treat fields with $B_0\lesssim 10^{-2}\,{\rm nG}$. Assuming $B_0= 10^{-2}\,{\rm nG}$, one thus expects $k_{\rm td}\simeq 2600-26000\,\Mpc^{-1}$. We will see that in fact the lower end of this range seems to be reproduced by the simulations (cf. Fig.~\ref{fig:mag_spectra_k09}).

\vspace{-3mm}
\subsubsection{Estimate for $z_{\rm td}$}
\label{sec:drag_analytic}
To determine the redshift $z_{\rm td}$ at which the turbulent decay resumes after recombination, we use the condition that the kinetic Reynolds number should be $R_{\rm e}\simeq 1$ at $k_{\rm fs}\approx k_{\rm td}$ and $z_{\rm td}$ and then $R_{\rm e} > 1$ afterwards. 
In the free streaming regime, we have $R_{\rm e}\approx V_{\rm bA}/\alpha_\gamma \ell_{\rm fs}\approx V_{\rm bA} R\, \ell_{\rm mfp}/\ell_{\rm fs}$, where the drag coefficient $\alpha_\gamma\approx  \Ne \sigT / R$ and $\ell_{\rm mfp}=1/\Ne \sigT$ for the mean free path were used \citep{Banerjee2004, Subramanian2016}. 
With $\ell_{\rm fs}=\lambda_{\rm fs} a = 2\pi a/k_{\rm fs}$, this yields\footnote{We use $R_{\rm e}^2=1$ to readily eliminate $B_0$ from the expression.} the condition
\begin{align}
\label{eq:def_td}
\int^\infty_{z_{\rm td}} \frac{\id z}{\Ne \sigT a H(z)}\approx \frac{2}{(n_B+5)}\,\left.\frac{R}{(2\pi\,\Ne \sigT a)^2}\right|_{\rm td},
\end{align}
for the redshift $z_{\rm td}$. We also define $k_{\rm t}= 2\pi \Ne \sigT a/\sqrt{R}$, such that Eq.~\eqref{eq:def_td} can be cast into the form $k_{\rm \lambda}(z_{\rm td})\approx \sqrt{(n_B+5)/2}\,k_{\rm t}(z_{\rm td})$.

In Fig.~\ref{fig:k_values}, we show a comparison of the different wave numbers relevant to the discussion.  The solution of Eq.~\eqref{eq:def_td} in our treatment here is independent of the magnetic field strength. For $n_B=-3$, we find $z_{\rm td} \approx 820$ and more generally
\begin{align}
\label{eq:z_td_approx}
z_{\rm td}\approx 820\left[1-\pot{4.66}{-2} \,(n_B+3)+\pot{1.39}{-3}(n_B+3)^2\right]
\end{align}
for $n_B\in[-3,3]$ to percent precision. This shows that in our treatment $z_{\rm td}$ drops gradually with increasing spectral index, $n_B$. We furthermore obtain
\begin{align}
\label{eq:k_td_lambda}
k_\lambda(z_{\rm td})&\approx \pot{7.50}{-3}\,\Mpc^{-1}\left[1-0.295 \,(n_B+3)
\right.
\nonumber\\&\quad+\left.\pot{4.66}{-2}(n_B+3)^2-\pot{2.97}{-3}(n_B+3)^2\right],
\end{align}
which determines $k_{\rm td}$ in Eq.~\eqref{def:kfs_rad_II}. This shows a moderate dependence of $k_{\rm td}$ on $n_B$, giving $k_{\rm td}\approx 20\,\Mpc^{-1}\left[B_0/1\,{\rm nG}\right]^{-1}$ for $n_B\simeq -3$ to $k_{\rm td}\approx 3.3\,\Mpc^{-1}\left[B_0/1\,{\rm nG}\right]^{-1}$ for $n_B\simeq 2$ in our treatment.
We will see below that both Eq.~\eqref{eq:z_td_approx} and Eq.~\eqref{eq:k_td_lambda} do not capture the dependence of the turbulent regime on $B_0$ and $n_{B}$, in particular when increasing $n_{B}$ (see Sect.~\ref{sec:var_B0} and Sect.~\ref{sec:var_nB}).

Comparing our results to previous works, a value $k_{\rm td}\approx 200-300\,\Mpc^{-1}\left[B_0/1\,{\rm nG}\right]^{-1}$ is usually quoted \citep[e.g.,][]{Sethi2005, Kunze2015, Subramanian2016}. Also, the redshift for turbulent decay usually is set to $z_{\rm td}\simeq 1100$ \citep[e.g.,][]{Sethi2005, Kunze2015}, independent of the PMF spectral index or strength. Here, we find lower values in both cases. The main reason for the different value at $n_{B}\simeq -3$ is the factor of $2\pi$ in the conversion from $\lambda_{\rm fs}$ to $k_{\rm fs}$. This yields the modified condition, $k_{\rm \lambda}(z_{\rm td})\approx k_{\rm t}(z_{\rm td})/2\pi$, for $z_{\rm td}$, which due to the exponential dependence of $k_{\rm \lambda}$ and $k_{\rm t}$ on redshift (see Fig.~\ref{fig:k_values}) results in $z_{\rm td}\simeq 1050$ instead. 
Overall, our simulations seem to suggest a scaling $k_{\rm td}\simeq 20\,\Mpc^{-1}\left[B_0/1\,{\rm nG}\right]^{-1}$ and $z_{\rm td}\simeq 800$ in agreement with our simple analytic estimate for $n_{B}=-3$.

\subsubsection{Magnetic field dissipation in the pre-recombination era}
\label{sec:pic_pre_rec}
The wavenumber $k_{\rm td}$ is only relevant if we wish to compare the initial power spectrum amplitude, $A$, for different values of $B_0$ and $n_B$. By construction, we already ensured that at $z=z_{\rm td}$ we have $\bar{B}^2 \approx B_0^2$ for all $n_B$. However, for different $n_B$ the effective damping scale at $z_{\rm td}$ differs, so that also the respective values for $A$ at the initial time vary. 
Consequently, we can also write
\begin{align}
\label{eq:B2_fs}
\bar{B}^2\approx B_0^2 \left(\frac{k_{\rm fs}(z)}{k_{\rm td}}\right)^{n_B+3}
=B_0^2 \left(\frac{k_{\lambda}(z)}{k_{\lambda}(z_{\rm td})}\right)^{2(n_B+3)/(n_B+5)}
\end{align}
at $z\gtrsim z_{\rm td}$, where we used Eq.~\eqref{def:kfs_rad_II} to eliminate $k_{\rm fs}(z)/k_{\rm td}$. We then have the magnetic dissipation rate\footnote{For this we need to compute $-a^{-4}\id (a^4 \rho_B)/\id t=-\rho_\gamma(z)\id ( \rho_B/\rho_\gamma)/\id t$.} at $z\gtrsim z_{\rm td}$
\begin{align}
\label{eq:QbarB_II}
\frac{\id (Q_{B}/\rho_\gamma)}{\id z}\approx \frac{2(n_B+3)}{n_B+5}\frac{\rho_B}{\rho_\gamma} \frac{\partial_z k_{\lambda}(z)}{k_{\lambda}(z)}
=\frac{(n_B+3)}{n_B+5}\, \frac{ k_{\lambda}^2(z) \,r_B(z)}{\Ne \sigT a H},
\end{align}
where we used $-(1/2) k^2_\lambda \partial_z k^{-2}_\lambda=k^{-1}_\lambda\partial_z k_\lambda$ and the magnetic energy density fraction $r_B(z)=\rho_B(z)/\rho_\gamma(z)$. 

In the radiation-dominated era, $\rho_B(z)\propto \rho_\gamma(z) \,k_\lambda^{2(n_B+3)/(n_B+5)}$ and $k_\lambda\approx \pot{1.8}{-6}(1+z)^{3/2}\,\Mpc^{-1}$ such that we find the redshift scaling $\id Q_{B}/\id z\propto \rho_\gamma(z) \,(1+z)^{2(n_B+2)/(n_B+5)}$. This is in excellent agreement with previous studies \citep{Banerjee2004, Wagstaff2015}, but differs slightly from \citet{Kunze2014}, who obtained $\id Q_{B}/\id z\propto \rho_\gamma(z) \,(1+z)^{(3 n_B+7)/2}$. The difference is because $k_{\rm fs}\propto \kD \propto (1+z)^{3/2}$ was used in their expression, which neglects the scaling of $k_{\rm fs}$ because of changes in $\rho_B$. For blue magnetic field power spectra, this overestimates the free-streaming heating, as also pointed out by \citet{Wagstaff2015}.

To compare with our numerical simulations, it is furthermore useful to give the scaling in the matter-dominated regime. 
Then $k_\lambda \propto (1+z)^{5/4}$ and $H\propto (1+z)^{3/2}$, such that one has $\id Q_{B}/\id z\propto \rho_\gamma(z) \,(1+z)^{(3n_B+5)/2(n_B+5)}$. 
This shows a slightly slower increase with redshift than for the radiation-dominated case. 
Our simulations reveal a scaling trend for magnetic dissipation which is similar and this is discussed in Sect.~\ref{sec:semi-analytics} with reference to Fig.~\ref{fig:heat_rate_comp_index}. 

\subsubsection{Creation of spectral distortions in the drag regime}
\label{sec:pic_pre_rec_dist}
In the drag-dominated phase, the dissipated magnetic field energy flows directly to the CMB photons, as these provide the viscosity in the free-streaming era. 
This causes a small $y$-distortion \citep{Jedamzik2000, Sethi2005, Wagstaff2015}, which physically is due to photon mixing \citep{Chluba2012}, and also sources CMB temperature anisotropies at ultra-small scales ($k \gtrsim 20\,\Mpc^{-1}\,[B_0/1\,{\rm nG}]^{-1}$). 
Since we are in the free-streaming regime, the photon mixing process is {\it incomplete} so that only a (small) faction of the magnetic field energy really creates a distortion. 
This is in stark contrast to the damping of primordial temperature fluctuation that erases all small-scale fluctuation due to photon diffusion, which allows $\simeq 1/3$ of the acoustic wave energy to create an all-sky average distortion \citep{Chluba2012}.

We also highlight that a small $y$-distortion has no significant effect on the baryon temperature and the recombination history as long as it remains below the {\it COBE/FIRAS} limit \citep{Chluba2008c}. 
Thus, this part of the PMF dissipation can only be constrained by measuring CMB spectral distortions and ultra small-scale CMB anisotropies, in spite of opposing recent claims \citep{Kunze2017}. 
Simple estimates, assuming that all the energy creates distortions, suggest a Compton $y$-parameter, $y\simeq 10^{-8}-10^{-7}$, for $B_0\simeq 1\,{\rm nG}$ \citep{Wagstaff2015}. 
Due to the finite resolution of the beam, which essentially leads to mixing of blackbodies of different temperature and hence a distortion \citep{Chluba2004}, these estimates provide useful upper limits; however, since many other sources of $y$-distortions at similar or larger level exist \citep[e.g.,][]{Refregier2000, Chluba2011therm, Hill2015}, it will be hard to use future distortion measurements to improve existing PMF limits in this way.

\begin{figure*}
 \centering 
    \begin{subfigure}[b]{0.49\textwidth}
 \includegraphics{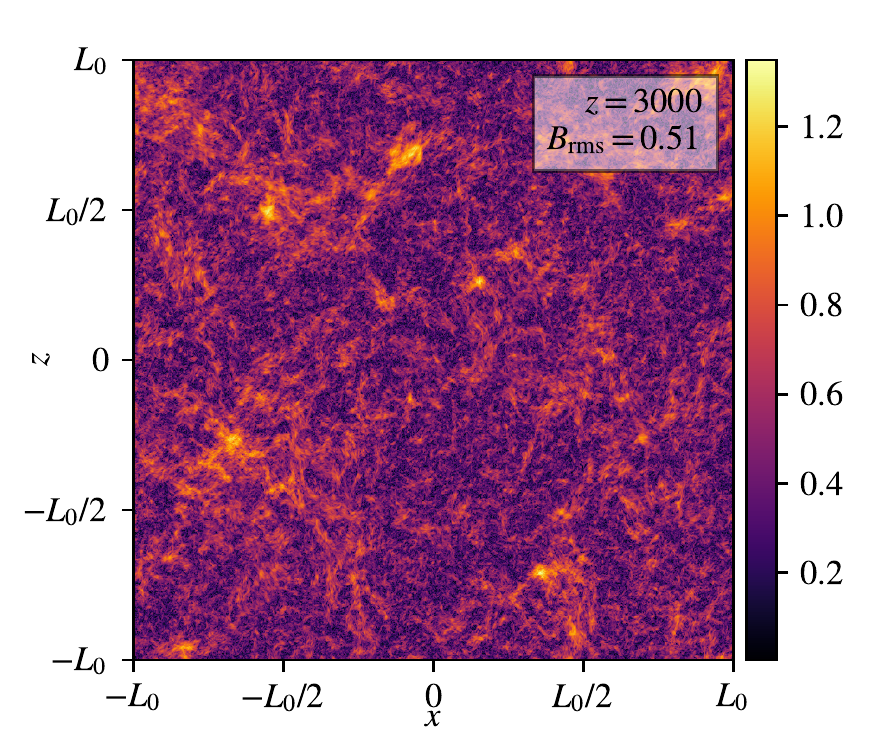}
 \label{fig:slice_bb_z3000}
 \end{subfigure}
\begin{subfigure}[b]{0.49\textwidth}
 \includegraphics{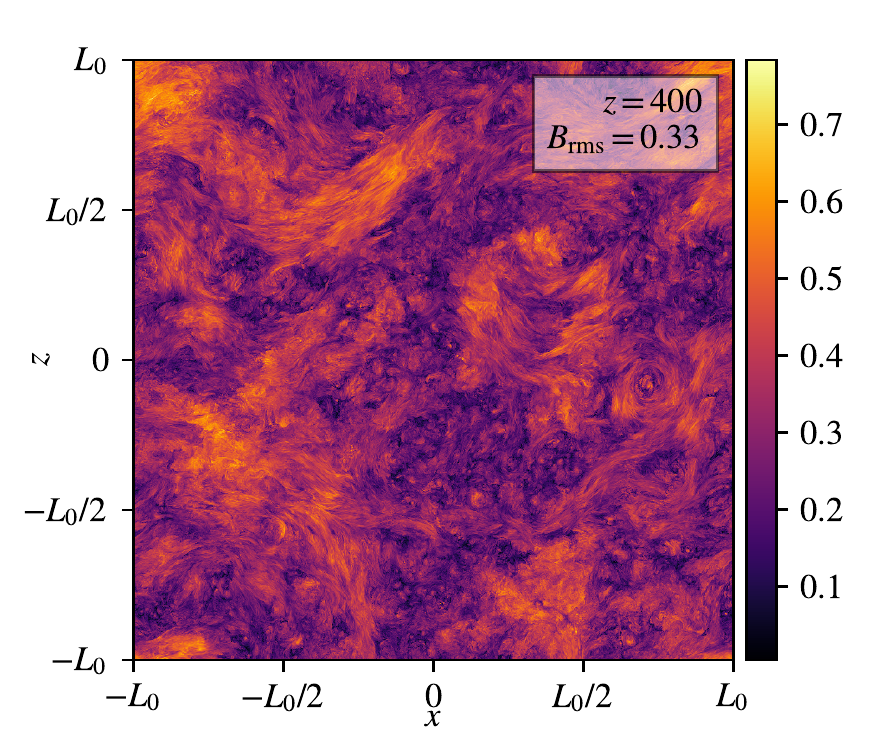}
 \label{fig:slice_bb_z400}
 \end{subfigure}
 \caption{Slices representing the magnetic field strength absolute value $|B|=\sqrt{B_x^2 + B_y^2 + B_z^2}$ through the $x-z$ plane at $y=0$ in the simulation box at two different redshifts, near the beginning of the simulation, at $z=3000$ (left panel) and near the end, at $z=400$ (right panel). The fields shown are for the fiducial case $n_B=-2.9$ and $B_0=0.51$ with both $|B|$ and $B_\mathrm{rms}$ expressed in code units of Alfv\'en velocity  divided by sound speed. Note the lower intensity scale for the evolved magnetic field strength at $z=400$.}
\label{fig:slices_b}
\end{figure*}

\subsection{Turbulent decay after recombination}
\label{sec:turb_decay}
In this section, we briefly recap the analytic approximations commonly used to describe the heating in the turbulent decay regime. 
We then provide an improved derivation that also includes the evolution of the magnetic field strength on the eddy turnover time. 
These approximations will be used later to represent the results from our numerical simulations (e.g., Sect.~\ref{sec:semi-analytics}).

\subsubsection{Review of previous analytical results}
\label{sec:turb_decay_old}
As argued by \citet{Banerjee2004}, in the post-recombination era turbulent decay dominates the damping of the magnetic field energy density with a evolution law 
\begin{align}
E_B(z)\approx E_{B, \rm td} \left[\frac{t}{t_{\rm eddy}}\right]^{-2(n_B+3)/(n_B+5)}
\end{align}
for $t\gg t_{\rm eddy}$. Here, $t_{\rm eddy}$ is the eddy turnover time of the largest turbulent mode at the initial time, $t_{\rm td}=t(z_{\rm td})$, and $E_B=|B|^2/8\pi=a^4 \rho_{B}$ is the corresponding comoving magnetic energy density. 

How do we generalize this to the expanding Universe? First of all, we want to obtain the limit $t\rightarrow t_{\rm td}$ for which $E_B(z)\rightarrow E_{B,\rm td}$. \citet{Sethi2005} suggested the generalization as
\begin{align}\label{eq:BK_sol}
E_B(z)\approx E_{B, \rm td} \left[1+\frac{t-t_{\rm td}}{t_{\rm eddy}}\right]^{-m}
\end{align}
with $m=2(n_B+3)/(n_B+5)$, which indeed has the right limit towards $t\rightarrow t_{\rm td}$. In the expanding Universe one also needs to replace
\begin{align}
\label{eq:t_tedd}
\frac{t-t_{\rm td}}{t_{\rm eddy}}\rightarrow \int \!\frac{\id t}{t_{\rm eddy}}
=\frac{1}{[t_{\rm eddy}H]_{\rm td}}\!\int_z^{z_{\rm td}} \!\frac{\id z}{1+z}
=\frac{1}{\tau_{\rm td}}\ln\left(\frac{1+z_{\rm td}}{1+z}\right).
\end{align}
Here, we have taken into account the explicit redshift dependence of the eddy turnover time, $t_{\rm eddy}(z)\simeq \ell_{\rm eddy}/\varv_{\rm bA}\approx t_{\rm eddy}(z_{\rm td}) \,(a/a_{\rm td})^{3/2}$, using $\varv_{\rm bA}\propto a^{-1/2}$ and $\ell_{\rm eddy} \propto a$, and assumed $H\propto a^{-3/2}$ as appropriate for the matter-dominated era. This neglects any evolution of $\varv_{\rm bA}$ related to $\rho_{B}$.
We also defined $\tau_{\rm td}=t_{\rm eddy}(z_{\rm td}) H(z_{\rm td})$, which is $\tau_{\rm td}\approx [R H/(\Ne\sigT c)]_{\rm td}\approx 0.063$ by construction. We thus obtain
\begin{align}
\label{eq:rho_turn_old}
E_B(z)\approx E_{B,\rm td} 
\left[1+\frac{1}{\tau_{\rm td}}\ln\left(\frac{1+z_{\rm td}}{1+z}\right)\right]^{-m},
\end{align}
which agrees well with previously used expressions \citep[cf.,][]{Sethi2005}. With this we find the dissipation rate at $z\lesssim z_{\rm td}$
\begin{align}
\label{eq:QbarB_III}
\frac{\id (Q_{B}/\rho_\gamma)}{\id z}\approx - \frac{1}{(1+z)} \frac{m}{\tau_{\rm td}} \,\frac{r_{B, \rm td} }{\left[1+\frac{1}{\tau_{\rm td}}\ln\left(\frac{1+z_{\rm td}}{1+z}\right)\right]^{m+1}},
\end{align}
which is usually used to describe the heating due to turbulent decay. Notice, that in contrast to the heating in the free-streaming regime, this term directly affects the baryons, raising their temperature and thus leading to changes of the cosmological recombination history. 

\subsubsection{Phenomenological derivation for the turbulent decay law}
The picture for decay of magnetic fields in the turbulent phase is that the time-scale on which energy is drained is given by the eddy turnover time around the critical scale that defines the comoving energy density of the fields, $E_B=|B|^2/8\pi=a^4\rho_B$. From Eq.~\eqref{eq:barB} it follows that $\lambda_{\rm c}\propto E_B^{-1/(n_B+3)}$. We then have
\begin{align}
\label{eq:eddy_rate}
\frac{1}{t_{\rm eddy}}\approx \frac{\varv_{\rm bA}}{\lambda_{\rm c} a}\approx\frac{1}{t_{\rm eddy, td}}\,
\left(\frac{E_B}{E_{B, \rm td}}\right)^{1/m}\,\left(\frac{a}{a_{\rm td}}\right)^{-3/2},
\end{align}
where we used $\varv_{\rm bA}\propto \sqrt{E_B/R}$ and set $m=2(n_B+3)/(n_B+5)$ as before. We normalized everything with respect to the values at $z_{\rm td}$. Then the evolution equation for the comoving magnetic energy density is given by
\begin{align}
\label{eq:dE_dt_eddy}
\frac{\id E_B}{\id t}&\approx -\frac{E_B}{t_{\rm eddy}}
=-\frac{E_B}{t_{\rm eddy, td}}
\left(\frac{E_B}{E_{B, \rm td}}\right)^{1/m}\left(\frac{a_{\rm td}}{a}\right)^{3/2}
=-\frac{E_{B}^{(m+1)/m}}{\tau_{\rm eddy, td} \,a^{3/2}}
\end{align}
with $\tau_{\rm eddy, td}=t_{\rm eddy, td}E_{B, \rm td}^{1/m}\,a_{\rm td}^{-3/2}$. Neglecting the expansion of the Universe, we can set $a=a_{\rm td}$ and then integrate from $t_{\rm td}$ to find
\begin{align}
\label{eq:Ev_sol}
E_B(t)\approx  E_{B, \rm td} \left[1+\frac{t-t_{\rm td}}{m t_{\rm eddy, td}}\right]^{-m},
\end{align}
which is only valid for $n_B>-3$. Comparing with Eq.~\eqref{eq:BK_sol}, shows that solution is a little different, with an extra factor $1/m$ modulating the eddy turnover time, which we will discuss below. 

Now including the effect of expansion during the matter-dominated era yields
\begin{align}
\label{eq:Eb_analytic}
E_B(t)\approx  E_{B, \rm td} \left[1+\frac{1}{m\tau_{\rm td}}\,\ln\left(\frac{1+z_{\rm td}}{1+z}\right)\right]^{-m}
\end{align}
in a similar manner. Comparing with Eq.~\eqref{eq:rho_turn_old}, we can see that again an extra factor of $1/m$ appears. 
Taking the redshift derivative yields
\begin{align}
\label{eq:QbarB_IV}
\frac{\id (Q_{B}/\rho_\gamma)}{\id z}\approx - \frac{1}{(1+z)} \,\frac{1}{\tau_{\rm td}}\,\frac{r_{B, \rm td} }{\left[1+\frac{1}{m\tau_{\rm td}}\ln\left(\frac{1+z_{\rm td}}{1+z}\right)\right]^{m+1}}
\end{align}
for the magnetic dissipation rate in the turbulent decay regime. As we can see, the heating amplitude differs by $1/m$ with respect to Eq.~\eqref{eq:QbarB_III} and scales slightly faster with redshift for near scale invariant initial spectra, since we have $m\rightarrow 0$ for $n_B\rightarrow -3$. 

Does the extra factor of $1/m$ in the evolution law make sense? For blue initial power spectra, the energy density is fully dominated by the smallest scales. For $n_B\rightarrow \infty $, we find $m\rightarrow 2$, so that the evolution law, Eq.~\eqref{eq:Ev_sol}, becomes
\begin{align}
E_B(t)\approx  E_{B, \rm td} \left[1+\frac{t}{2 t_{\rm eddy, td}}\right]^{-m}.
\end{align}
This indicates that the relevant time for energy dissipation is about twice the eddy turnaround time, a result that is also found in typical turbulent flow simulations. 
For lower values of $n_B$, the evolution affects the total energy density of the magnetic fields to a smaller extent and the eddy turnover time also scales less strongly with time. In the limit of quasi-constant strength of fields at all scales ($n_B\approx -3$), we also have $t_{\rm eddy}(E_B)\approx {\rm const}$, which means that the field strength decays exponentially at a rate $1/t_{\rm eddy}$. This limit is more problematic, since in this case the true small-scale dissipation time-scale, $t_{\rm turb}$, will determine the rate of change of the energy density and not the time-scale at which energy is injected into the turbulent cascade ($\simeq t_{\rm eddy}$). However, for $n_B>-3$ our approximate model should still provide a reasonable description of the problem in the different regimes, as long as $ t_{\rm eddy}\gg t_{\rm turb}$.

A last argument why the factor $1/m$ seems appropriate is that we know that at the initial redshift $z_{\rm td}$ we should find 
\begin{align}
\frac{\id (Q_{B}/\rho_\gamma)_{\rm td}}{\id z} 
&\equiv \frac{1}{[H (1+z)]_{\rm td}}\frac{\id a^4 \rho_B(z_{\rm td})}{a^4\id t} 
\approx  -\frac{1}{[H (1+z)]_{\rm td}}\, \frac{r_{B, \rm td}}{t_{\rm eddy, td}} 
\nonumber
\\
&=-\frac{r_{B, \rm td}}{(1+z_{\rm td})\tau_{\rm td}}, 
\end{align}
as is directly obtained from Eq.~\eqref{eq:QbarB_IV} for $z=z_{\rm td}$ but is {\it not} reproduced by the law in Eq.~\eqref{eq:QbarB_III}. 
This shows that the magnetic dissipation rate at the transition from free-streaming damping to turbulent decay for fixed $\rho_B(z_{\rm td})$ is independent of $n_B$.

We will use Eq.~\eqref{eq:QbarB_IV} as a starting point for analytic representations of our numerical results. Even if it does not capture the physics completely, it does represent the late time evolution very well (see Fig.~\ref{fig:heat_rate_comp_index}). Together with Eq.~\eqref{eq:QbarB_II} for the pre-recombination evolution, we obtain a very useful semi-analytic description over a wide range of PMF parameters (also see Sect.~\ref{sec:semi-analytics}).

\begin{figure}

    \begin{subfigure}[b]{.49\textwidth}
 \includegraphics{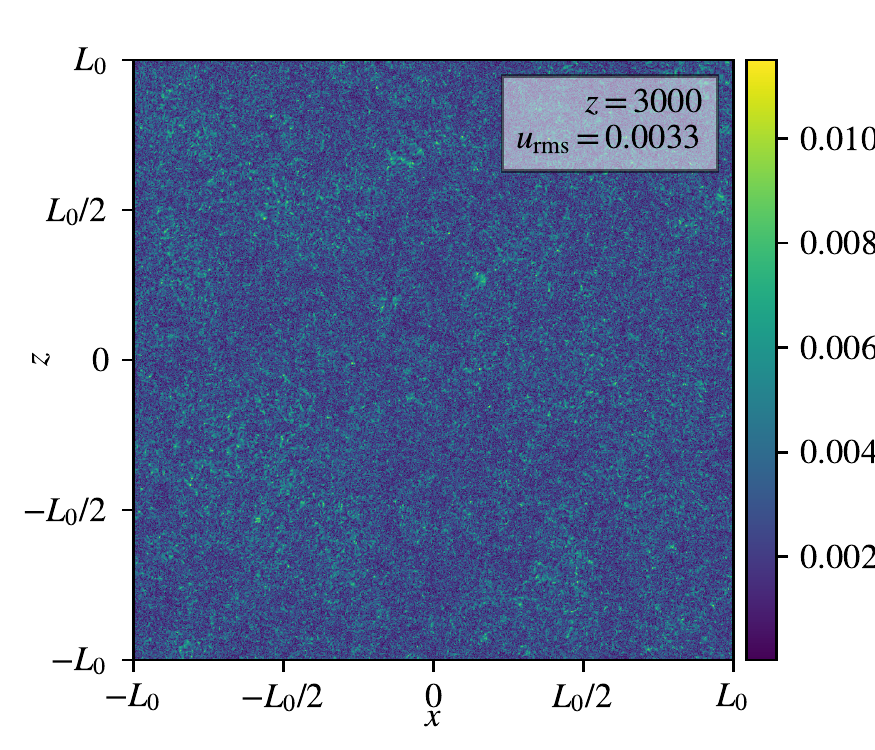}
 \label{fig:slice_uu_z3000}
 \end{subfigure}
\begin{subfigure}[b]{.49\textwidth}
 \includegraphics{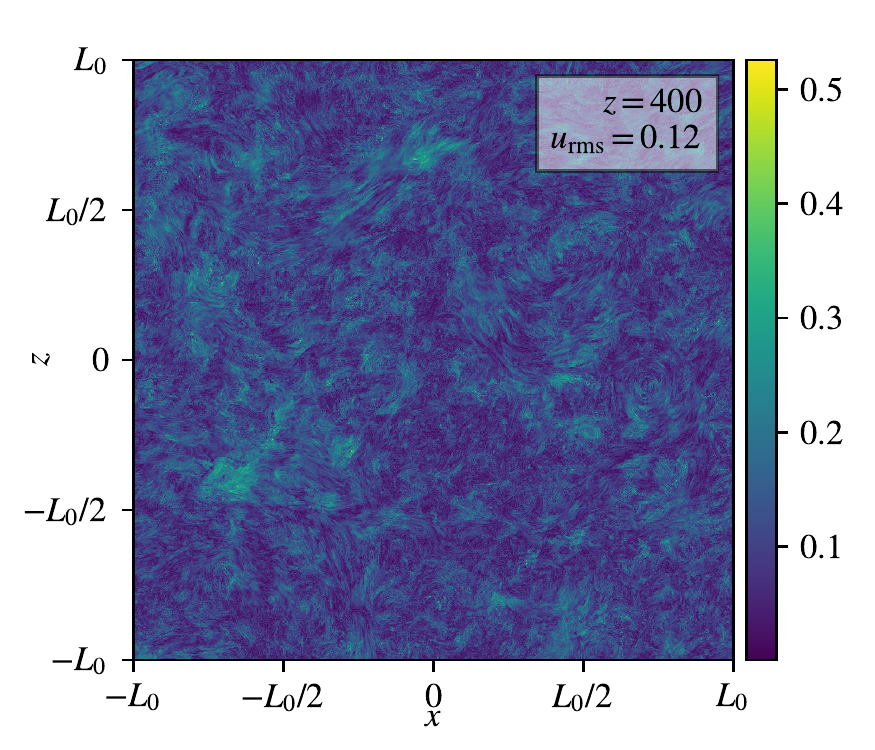}
 \label{fig:slice_uu_z400}
 \end{subfigure}
 \caption{Slices representing the baryon velocity absolute value $|u|=\sqrt{u_x^2 + u_y^2 + u_z^2}$ through the $x-z$ plane at $y=0$ in the simulation box at two different redshifts, $z=3000$ (upper panel) and $z=400$ (lower panel). Both $|u|$ and $u_\mathrm{rms}$ are in code units (i.e. divided by sound speed). Note the 50 times smaller intensity scale for $|u|$ at $z=3000$.}
\label{fig:slices_u}
\end{figure}

\vspace{-2mm}
\section{Simulations of Heating by primordial magnetic fields}
\label{sec:simulations}
In this section, we describe the setup of our numerical magnetohydrodynamic (MHD) simulations to follow the evolution of cosmological magnetic and velocity fields from the photon drag dominated regime, across the recombination era, into the turbulent decay dominated regime. Our goal is to compute and follow the evolution of the net heating produced by the dissipation of magnetic energy. 
We first detail the numerical methods, the code employed, the form of the MHD equations and the integration scheme used. The choice of cosmological coordinates and initial conditions is then described and the evolution of the photon drag is presented. Conversion to physical units is also discussed. We then briefly discuss how to use the results to extract different dissipation and heating rates. 

\vspace{-2mm}
\subsection{Numerical Method and Setup}
\subsubsection{The Pencil-Code}
We perform full 3-D simulations with the \texttt{Pencil-Code} \citep{PENCIL10}, at a resolution of $N=1536^3$, to numerically investigate the details of heating and sourcing of turbulence by magnetic fields in the context of varying photon viscosity across the epoch of recombination. Two-dimensional slices of the simulated 3D magnetic and baryon velocity fields, near the start and end epochs, are shown in Figs.~\ref{fig:slices_b} $\&$ \ref{fig:slices_u} and discussed in section~\ref{subsec:fiducial_u_b}.

The \texttt{Pencil-Code} is a finite-differences code with a sixth-order integration scheme. It solves the equations that govern the flow of the plasma and its interactions with the magnetic field for every grid point individually using the values of three neighbouring cells for the derivatives. This method means we are limited to subsonic velocities and sub-sonic Alfv\'en velocities, which limits us to a physical magnetic field strength $B_{\rm phys}\lesssim \text{few}\times10^{-2}\,{\rm nG}$ (Sect.~\ref{sec:phy_units}). In general, the density fluctuations are of the order $\simeq 1\%$, corresponding to (nearly) incompressible flows. 

The code integrates the MHD equations in the following form,
\vspace{-2mm}
\bsub
\label{eq:MHD}
\begin{align}
    \frac{\mathrm{D}\ln\rho}{\mathrm{D} t} &= -\nabla \cdot \mathbf{u} \label{eq:MHD1} 
    \\
    \frac{\mathrm{D}\mathbf{u}}{\mathrm{D}t} \,\,\,&= -\frac{\nabla p}{\rho} + \frac{\mathbf{j}\times \mathbf{B}}{\rho} + \mathbf{f}_\mathrm{visc}  \label{eq:MHD2}
    \\
    \frac{\partial\mathbf{A}}{\partial t} \,\,\,&=  \,\,\,\mathbf{u}\times \mathbf{B} - \eta\ \mathbf{j} \label{eq:MHD3},
\end{align}
\esub
where $\mathrm{D}/\mathrm{D}t= \partial /\partial t + \mathbf{u}\cdot \nabla$ is the convective derivative, $\rho$ describes the (mass) density, $\mathbf{u}$ the velocity field of the plasma and $\mathbf{j}=\nabla\times \mathbf{B}$ is the MHD current in the simulation. Equation~(\ref{eq:MHD1}) is the continuity equation of the plasma, Eq.~(\ref{eq:MHD2}) is the Navier-Stokes equation, and Eq.~(\ref{eq:MHD3}) is the induction equation of the magnetic field.

There also is a magnetic resistivity $\eta$ in the code, meaning we are working with \textit{non-ideal} MHD. It is set in such a way that the magnetic Prandtl number, the ratio of viscosity to resistivity, is given by $\mathrm{Pr}=\nu_3/\eta_3=1$. The code uses the vector potential $\mathbf{A}$, with $\mathbf{B}=\nabla\times\mathbf{A}$ as a primitive variable, as it naturally ensures the divergence free condition $\nabla\cdot \mathbf{B}$ for the magnetic field. 

We do not explicitly solve the energy equation in our simulation, since we use a isothermal equation of state where the pressure is only dependent on the density in the simulation with $p=c_s^2\rho$.
Assuming a mono-atomic gas of particles, the sound speed is then given \citep[e.g.][]{Ma1995} by
$c_s= \sqrt{4/3\ k_\mathrm{B}\ T_\mathrm{b}/m_p}\approx 5.7\,{\rm km\,s^{-1}}\,\left[(1+z)/1100\right]^{1/2}$.
We also assume a post-recombination neutral hydrogen gas in thermal equilibrium with the CMB photon field, $T_{\rm b}\simeq T_\gamma\propto a^{-1}$, which is valid until $z\lesssim 150$. 
At scales below the photon diffusion scale (most relevant to our problem), the expression employed for the sound speed is valid even in the pre-recombination era, as photons and baryons no longer behave as a single tightly-coupled fluid ($c_s = c/\sqrt{3}$) in the free-streaming regime \citep[see][for a more detailed discussion]{Jedamzik1998, Banerjee2004}. 
At larger scales, corrections could become relevant, but the effect of damping is deemed negligible at those scales.
We also neglect the effect of recombination on the sound speed, which is related to the change in the number of free particles. 
For instance, for a fully ionized electron-proton plasma, one would have\footnote{The more general expression should read $c_s^2\approx\frac{4}{3}(1-Y_{\rm p})(1+f_{\rm He}+X_{\rm e})\frac{k\,T_{\rm b}}{m_p}$, where $f_{\rm He}\approx 0.079$ for $Y_{\rm p}=0.24$ and $T_{\rm b}\approx T_\gamma\propto a^{-1}$.} $c^{\rm ep}_s\approx\sqrt{2}c_s\simeq 1.4 \,c_s$. 

The last term in Eq.~(\ref{eq:MHD2}) describes the viscosity of the baryon-photon fluid. It includes the effect of photon drag, which we model as a viscosity acting on the fluid, and additional numerical viscosity to avoid numerical instabilities, as well as to mimic sub-grid dissipation effects in the fully turbulent regime:
\begin{align}
    \mathbf{f}_\mathrm{visc} &= \textbf{f}_\mathrm{hyper} + \textbf{f}_\mathrm{drag} \label{eq:visc_tot}\\
    \mathbf{f}_\mathrm{hyper} &= \nu_3\nabla^6 \mathbf{u} \label{eq:visc_num}\ \\
    \mathbf{f}_\mathrm{drag} &= -\alpha(z) \cdot \mathbf{u} \label{eq:visc_drag}.
\end{align}
The effect of numerical viscosity $\mathbf{f}_\mathrm{hyper}$ is controlled by the variable $\nu_3$, which we optimize to avoid large pile-up of power at small-scales in the turbulent phase (see Sect.~\ref{sec:num_vis} and App.~\ref{sec:appendix1} for a detailed discussion of convergence w.r.t. numerical viscosity in our simulations). 
We discuss the implementation of the photon drag term $\alpha(z)$ in Sect.~\ref{sec:photon_drag}. 
Our simulations neglect all gravitational interactions of the plasma as well as any contributions from photon perturbations.

\subsubsection{Transformation to Cosmological Coordinates}
The {\tt Pencil}-Code works in dimensionless code units, with time expressed in units of the Hubble time, assuming the matter-dominated regime, $t_{\rm exp}=H^{-1}= t_0\,a^{3/2}\approx \pot {8.2}{17} (h/0.7)^{-1}\,a^{3/2}\,{\rm s}$. Here, we do not explicitly treat the radiation-dominated regime, which only changes the redshift-dependence of the drag-dominated dissipation rate (see Sect.~\ref{sec:pic_pre_rec}).
We furthermore use the {\it super-comoving} coordinates \citep{Banerjee2004}:
\bsub
\begin{align}
&\tilde{\rho}_{\rm b}= a^3 \rho_{\rm b}  
&\tilde{p}= a^4 p &
&\tilde{\vek{B}}=a^2 \vek{B} &
&\tilde{\vek{u}}=a^{1/2} \vek{u} &
\\
&\id \tilde{t}= \id t/t_{\rm exp}
&\tilde{\eta}= a^{-1/2} \eta &
&\tilde{\vek{A}}=a \vek{A} &
&\tilde{\vek{j}}=a^{7/2} \vek{j} &.
\end{align}
\esub
This coordinate transformation is a rescaling that uses the conformal property of the MHD equations allowing us to describe MHD in an expanding FRW Universe via the form of the MHD equations in a flat spacetime (see Eq.~\ref{eq:MHD}). 
Only one additional term is introduced, a kinematic Hubble drag described in the next sub-section.
Within the simulation, the comoving sound speed $\tilde{c}_s=1$, so that computational velocities or magnetic field amplitudes can be converted to physical units using $\vek{\varv} = c_s \tilde{\vek{u}}$ and $\vek{B} = \sqrt{4\pi\,\rho_{\rm b}}\, c_s \,\tilde{\vek{u}}_{\rm A}$, where $\tilde{\vek{u}}_{\rm A}$ is the Alfv\'en velocity obtained directly from the code. 
To map from $\tilde{t}$ to redshift, we use $\id \tilde{t}=\id t/t_{\rm exp}\equiv -\id \ln (1+z)$, where $\id t = -t_{\rm exp} a \id z$. We obtain $\tilde{t}=\ln[(1+z_0)/(1+z)]$, where $z_0$ is the starting redshift. This implies $(1+z)=(1+z_0)\,\expf{-\tilde{t}}$, which is then used to compute the evolving photon drag coefficient $\alpha(z)$.

\begin{figure}
    \includegraphics{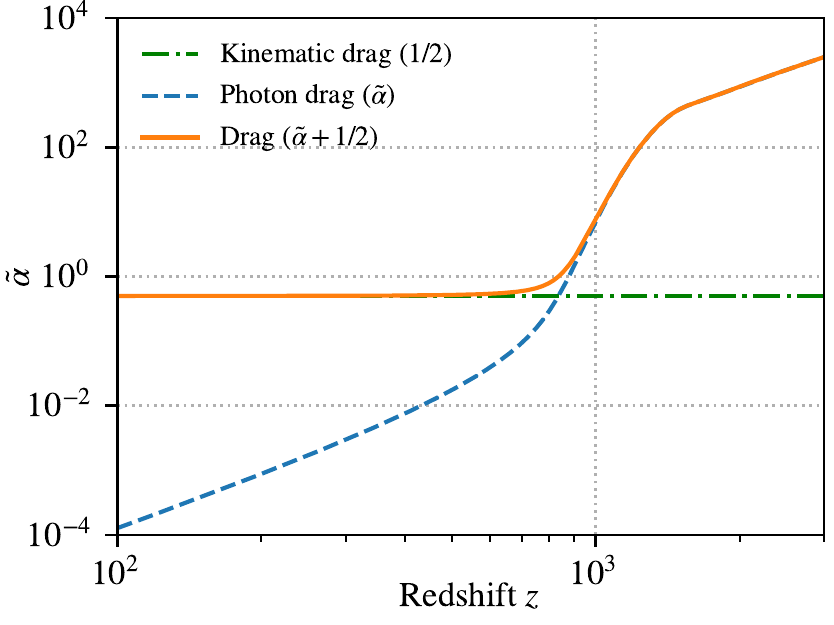}
    \caption{The evolution of the photon drag coefficient $\tilde{\alpha}$, Eq.~\eqref{eq:alpha_drag}, in the simulation code units (where rates are in units of Hubble rate) as well as the constant kinematic drag coefficient $H/2$ (or 1/2 in code units), which is important only at late epochs.}
    \label{fig:effective_rate_k09}
\end{figure}

\subsubsection{Photon Drag}
\label{sec:photon_drag}
The occasional scattering of photons off electrons in the free-streaming regime damps magnetic field power at small scales \citep{Jedamzik1998}. 
In physical units, the drag coefficient is given by \citep[e.g.,][]{Banerjee2004}
\begin{align}
\alpha &= \frac{c}{R \ell_{\rm mfp}}=\frac{\sigT\Ne c}{R}=\frac{\sigT N_{\rm H}c}{R}\,X_{\rm e},
\end{align}
where $N_\mathrm{H}\approx \pot{1.9}{-7}\,(1+z)^3\,{\rm cm^{-3}}$ is the hydrogen nuclei number density\footnote{Assuming helium fraction $Y_{\rm p}=0.24$.} and $X_{\rm e}$ the free electron fraction, which is provided by \texttt{CosmoRec} \citep{Chluba2010b}. 
The time transformation $\id t \rightarrow \id \tilde{t}$, implies that the rescaled drag coefficient is $\tilde{\alpha}=t_{\rm exp}\alpha$, relevant in the computation. Inserting numerical values then yields
\begin{align}
\label{eq:alpha_drag}
\tilde{\alpha} 
&\approx \pot{4.6}{-6}
\,\left[\frac{\Omega_{\rm m} h^2}{0.14}\right]^{-1/2}\left[\frac{T_0}{2.726\,\Kel}\right]^{4}\,X_{\rm e}\,\,(1+z)^{5/2},
\end{align}
where $\Omega_{\rm m} h^2$ sets the total matter density. The evolution of $\tilde{\alpha}$ is illustrated in Fig.~\ref{fig:effective_rate_k09}. At $z\gg 1100$, the photon drag is dominant and prevents turbulence from developing. As recombination proceeds, the drag coefficient drops quickly and the flow becomes turbulent (also see discussion of the magnetic Reynolds number cf. Fig.~\ref{fig:reynold_amplaa_comp}). 

In the final form of the velocity equation~\eqref{eq:MHD2}, an extra kinematic drag term $-\frac{1}{2} H \vek{u}$ appears due to our choice of scalings corresponding to an expanding Universe \citep[cf.][]{Banerjee2004}. This term is usually neglected; however, it becomes significant after recombination (see Fig.~\ref{fig:effective_rate_k09}) and can be incorporated by replacing $\tilde{\alpha}\rightarrow \tilde{\alpha} + \frac{1}{2}$ inside the code. In the computation of the net heating rate, this term has to be taken into account as it reduces the net heating rate noticeably.

\vspace{-2mm}
\subsubsection{Physical Units}
\label{sec:phy_units}
To convert our code units into physical units, we use the physical sound speed $c_s \approx 5.7\,{\rm km\,s^{-1}}\,\left[(1+z)/{1100}\right])^{1/2}$ at recombination, evaluated at the current epoch $ c_s \approx 0.17\, {\rm km\,s^{-1}}\,\left({1+z}\right)^{1/2} $ as a comoving sound speed. 
We assume $\Te=\Tg\propto (1+z)$ at all times. This approximation is valid in the standard recombination scenario until $z\simeq 150$; however, significant heating by magnetic fields could change this, an effect that we do not treat in our simulations.

We can then determine the comoving physical box size using the distance an Alfv\'en wave travels within a Hubble time $t_0$. 
Taking the Alfv\'en speed equal to the plasma sound speed gives a box size $L_{\rm phys} \approx t_0 c_s L_\mathrm{code} \approx 28 \, (h/0.7)^{-1}\,{\rm kpc}$, where $L_\mathrm{code}=2\pi$ for our simulation box. 
The wavenumber in physical units then can be calculated from the box size, $L_{\rm phys} \approx 2\pi/k_{\rm phys}$ yielding $k_{\rm phys} \approx 220\, (h/0.7)$ Mpc$^{-1}$. 
For the smallest resolved physical scale, we therefore have $k_{\rm res}\simeq (N/2)\,k_{\rm phys}\simeq \pot{1.7}{5}\, (h/0.7)$ Mpc$^{-1}$ with a typical number of grid points $N = 1536$.

The magnetic field strength from an Alfv\'en velocity equivalent to the sound speed is given by $B_{\rm phys} = \sqrt{4 \pi \,\rho_{\rm b}} \,c_s \approx 3.9 \times 10^{-2}$~nG. 
To avoid the requirement of including significant density perturbations and shocks, we must limit Alfv\'en velocities to $\varv_{\rm bA} \lesssim c_s/2$. 
Thus, for the simulations considered here (and presented in the subsequent figures) the comoving physical magnetic field strengths lie in the range $B_{\rm 0, phys} \simeq (0.086-2.0) \,\times \,10^{-2}$ nG, corresponding to Alfv\'en velocities in the range $\simeq (0.022-0.51)\,c_s$.

\subsubsection{Initial Conditions}
In our simulations, we initialize the magnetic field with a specific power spectrum $P_M \propto k^n$ with a spectral index $n$, recalling that $n=n_\mathrm{B}+2$. 
This is done by initializing a Gaussian random field in $k$-space which is then multiplied with the chosen power-law and transformed back into real space. 
By doing this we also have full control over the helicity of the magnetic field $\mathcal{H}=\mathbf{A}\cdot \mathbf{B}$, which we set to zero at the initial epoch, focusing on non-helical fields. 
Since magnetic helicity is a conserved quantity (although strictly, only in ideal MHD) this remains constant for the entire simulation, except for numerical fluctuations. 
By choosing the amplitude of the magnetic power spectrum we can fix the $B_\mathrm{rms}$ in our simulation. 
This is also discussed in Sect.~\ref{sec:av_B} (where magnetic power spectrum was denoted as $P_B(k)$ and the spectral index $n_\mathrm{B}$ differs by two: $n =n_\mathrm{B}+2$). 

The velocity field is initialized with $\mathbf{u}\mid_{t=0}=0$. After a few time steps velocity fluctuations of the order $u_\mathrm{max}\approx c_s$ develop, which then damp out. 
This \textit{ring-in} phase is usually restricted to the high redshifts $z \sim 5000$, but for very steep spectra or low magnetic field strengths, it can last until $z\simeq 3000$. 
We show our results and figures for times $z \le 3000$, after this phase has ended and the simulation has gone into a near steady state. 
Unless stated otherwise, we keep track of the evolution of the magnetic and velocity field during the initial phase and rescale our curves to their values at $z=3000$, taking that epoch as the relevant initial condition for physical interpretation. 

\vspace{-2mm}
\subsubsection{Numerical viscosity and real heating}
\label{sec:num_vis}
We use the hyperviscosity implementation of the \texttt{Pencil-Code} since it gives the lowest numerical dissipation and largest Reynolds numbers in the simulation. 
A small amount of numerical viscosity is required but we do not want it to be stronger than the physical effects we wish to observe and follow. 
For a given resolution one can use a specific minimum viscosity pre-factor $\nu_3$. 
To further reduce this pre-factor, one has to increase the resolution $N$. 
For a given viscosity, $\nu_3$, the simulation is independent of resolution as long as one is above the minimum value of the viscosity, which is a function of resolution. 
On the other hand there must exist enough dissipation when the photon drag becomes negligible. 
A detailed discussion of how the viscosity pre-factor was chosen as well as the convergence tests of resolution can be found in App.~\ref{sec:appendix1}. 
The adopted value of $\nu_3=\pot{2.5}{-16}$ is used in all the studies below where when no explicit viscosity parameter values are mentioned.

\vspace{3mm}
\subsection{Theoretical considerations for magnetic heating rates}
In our simulations, the PMFs are the origin of energy that can subsequently be dissipated. 
Physically, the PMFs drive baryon velocities through the Lorentz force, sourcing a kinetic energy density, $\rho_{\rm kin}=\frac{1}{2}\,\rho_{\rm b} \left<\varv_{\rm b}^2\right>$. 
This process builds up a flow, which itself loses energy through i) interactions with CMB photons (via photon drag) and ii) dissipative processes at ultra-small scales (via Coulomb interactions and plasma effects). 
Photon drag leads to small-scale perturbations in the photon temperature, which through photon mixing processes can cause $y$-type distortions (see Sect.~\ref{sec:pic_pre_rec_dist}). 
The dissipative processes at ultra-small scales lead to heating of the baryons, which becomes important in the turbulent phase of the evolution.

Neglecting dissipation processes, the total (comoving) energy density\footnote{The photon energy density is not affected significantly and will be assumed to remain unaltered throughout the evolution.}, $E_{\rm tot}=E_B+E_{\rm kin}$, related to magnetic fields is conserved. 
Here, $E_B=a^4 \rho_B\equiv \frac{1}{2}\,a^4 \rho_{\rm b}\left<\varv^2_{\rm A,b}\right>$ and $E_{\rm kin}=a^4 \rho_{\rm kin}=\frac{1}{2}\,a^4 \rho_{\rm b}\left<\varv^2_{\rm b}\right>$. 
Real changes to the total energy are then given by
\begin{align}
\frac{\id E_{\rm tot}}{\id t}
&=\frac{\id E_B}{\id t}+\frac{\id E_{\rm kin}}{\id t}=-2\alpha E_{\rm kin}- H E_{\rm kin} - \left.\frac{\id E_{\rm tot}}{\id t}\right|_{\rm heat}.
\end{align}
The first term on the r.h.s. is due to photon drag, which can be obtained as $\dot{E}_{\rm kin}|_{\rm drag}\simeq \frac{1}{2}\,a^4 \rho_{\rm b}\left<2\,\varv_{\rm b}\,\dot{\varv}_{\rm b}\right>\equiv-2\alpha\,E_{\rm kin}$ using $\dot{\varv}_{\rm b}=-\alpha \varv_{\rm b}$.
The second term is related to the extra cooling from Hubble expansion. This term causes extra losses from the velocity field and has to be eliminated when computing the net heating rates.
The last term is due to dissipative processes in the baryon plasma at ultra-small scales, which leads to real heating.

Around the recombination era, three phases exist: at redshift $z\gg z_{\rm td}$, the dissipation is dominated by photon drag, so that 
\begin{align}
\frac{\id E_{\rm tot}}{\id t} &\approx -2\alpha E_{\rm kin}.
\end{align}
As is clear from Fig.~\ref{fig:effective_rate_k09}, the Hubble term can be neglected here.
In this phase, {\it no real heating occurs}, but perturbations in the photon fluid are generated at small scales through the Doppler effect. 
In this regime the drag is so strong that only very small velocity fluctuations are generated by magnetic fields (via the Lorentz force). The drag is strong enough to prevent magnetic fields from accelerating baryons up to the Alfv\'en-velocity, such that turbulence is highly suppressed (see Sect.~\ref{sec:Results} and Fig.~\ref{fig:tser_drag_k09}).

Around $z\simeq z_{\rm td}\simeq 10^3$, hydrogen atoms form and photon drag drops rapidly. 
The velocity field slowly builds up until a turbulent flow is formed. However, in the transition phase, little energy is actually dissipated, and most of the energy lost by the PMFs is dumped to increase the flow's kinetic energy (see Fig.~\ref{fig:tser_drag_k09}). 

In the turbulent phase ($z\lesssim z_{\rm td}$), photon drag can be neglected and a significant fraction of the energy lost from the magnetic fields is converted into heat by the turbulent cascade:
\begin{align}
\frac{\id E_{\rm tot}}{\id t} & \approx - H E_{\rm kin} - \left.\frac{\id E_{\rm tot}}{\id t}\right|_{\rm heat}.
\end{align}
In the simulations, we can thus compute the real heating of the medium in the various phases as
\begin{align}
\label{eq:net_heating_rate_def}
\left.\frac{\id E_{\rm tot}}{\id t}\right|_{\rm heat}\approx -\frac{\id E_B}{\id t}-\frac{\id E_{\rm kin}}{\id t}- H E_{\rm kin} -2\alpha E_{\rm kin}.
\end{align}
We will see below that the above picture is reproduced by the simulations. In particular, the transition phase between the drag-dominated and the turbulent regime is a new physical regime that was previously missing or treated as unrealistically abrupt.

\vspace{-3mm}
\subsubsection{Extraction of heating rates from simulation}
Within our simulations, the scales responsible for real heating of the medium (due to Coulomb interactions) are not resolved but instead mimicked by numerical viscosity to reproduce a quasi-Kolmogorov turbulent cascade with magnetic field power spectrum similar to $E_k\propto k^{-5/3}$ (see Sect.~\ref{sec:num_vis}). 
To compute the total energy loss rate of the PMFs, we can use the definition of the Alfv\'en speed, $V_{\rm bA}$, Eq.~\eqref{def:Alfven_speed_etc} for the ratio of the average magnetic to photon energy density. 
This yields
\begin{align}
\label{eq:r_B}
\left<r_B\right>=\frac{\left<\rho_B \right>}{\rho_\gamma}=\frac{2 R}{3} \, \left<V_{\rm bA}^2\right>
=\frac{2 R}{3} \, \frac{c^2_{\rm s}}{c^2}\,\left<\tilde{u}^2_A\right>,
\end{align}
where we used $V_{\rm bA}^2=(c^2_{\rm s}/c^2) \, \tilde{u}^2_A$. We recall that $\tilde{u}_A$ is the Alfv\'en velocity obtained directly from the code. Noting that $R \propto a$ and $c^2_{\rm s} \propto a^{-1}$, we can then compute the dissipation and heating rates in physical units in terms of the rate of change of Alfv\'en velocity $\tilde{u}_A$ or magnetic field amplitude $B_0$, both squared, and both in the same code units,
\begin{align}
\label{eq:Q_{B comp}}
\frac{\id (Q_B/\rho_\gamma)}{\id \ln z}
=-\frac{\id \left<r_B\right>}{\id \ln z}
&\approx - \pot{1.4}{-10}\frac{\id \quad \,\,}{\id \ln z} \left<\tilde{u}^2_A\right> \nonumber \\
&\approx - \pot{3.7}{-11}\frac{\id \quad \,\,}{\id \ln z} \left(\left<\frac{B_0}{0.51}\right)^2\right>,
\end{align}
where $c^2_{\rm s}/c^2\simeq \pot{3.2}{-13}(1+z)$ and the value $B_0=0.51$ corresponds to our fiducial case.
 Note that the magnetic dissipation rate ${\rm d}Q_B/{\rm d}\ln z$ or ${\rm d}E_{\rm mag}/{\rm d}\ln z$ in Fig.~\ref{fig:heat_rate_k09} is the negative rate of change of magnetic energy density ${\rm d}\langle \rho_B\rangle /{\rm d}\ln z$. 
Similarly, we can define the changes to the kinetic energy density of the fluid and net heating rates through Eq.~\eqref{eq:net_heating_rate_def}.

\section{Simulation Results for different PMF parameters}
\label{sec:Results}
The subsequent sections describe our results for the evolution of the amplitudes and power spectra of both magnetic and velocity fields as well as for the derived dissipation and heating rates. Initially, we present simulation results for a single value of magnetic field strength, chosen with a near scale-invariant magnetic power spectrum (Sect.~\ref{sec:scale-invariant}). Subsequently, we discuss results for varying magnetic field amplitude as well as varying magnetic spectral index (Sect.~\ref{sec:var_B0} and \ref{sec:var_nB}). We also compare our results with analytical expressions, highlighting expected trends and differences, providing analytic fits where relevant. 

\vspace{-0mm}
\subsection{Scale-invariant magnetic spectrum}
\label{sec:scale-invariant}
We first present simulation results for a single fiducial case with magnetic field amplitude $B_0 = 0.51$ (in code units of $\varv_A/c_s$) at $z=3000$. This field amplitude is equivalent to $B_{0, \rm phys} \approx 2.0 \times 10^{-2}$ nG. This fiducial case is chosen for a nearly scale-invariant power spectrum $n_B = -2.9$ (or alternatively, $n = n_\mathrm{B} + 2 = -0.9$), which is of particular importance in the study of primordial magnetic fields generated by inflation.

\vspace{-0mm}
\subsubsection{Evolution of magnetic and kinetic amplitude}
\label{subsec:fiducial_u_b}
A visual representation of the changing magnetic field strength and baryon velocity over 2D slices from our 3D simulations is shown in Figs.~\ref{fig:slices_b} $\&$ \ref{fig:slices_u}. 
A video of the temporal evolution of both the magnetic and kinetic field intensity slices can also be viewed at \url{https://www.hs.uni-hamburg.de/research/cosmo-mf}. 
Note the varying intensity scale between $z=3000$ $z=400$, particularly for the velocity in Fig.~\ref{fig:slices_u}. 
We can clearly see how the magnetic field decays slowly before recombination, evolves quicker over the transition regime and then spatially mixes and decays in the turbulent regime. 
The velocity field is boosted from zero initial value to saturation toward the end of the transition regime and subsequently exhibits turbulent mixing and decay. 

\begin{figure}
   \includegraphics{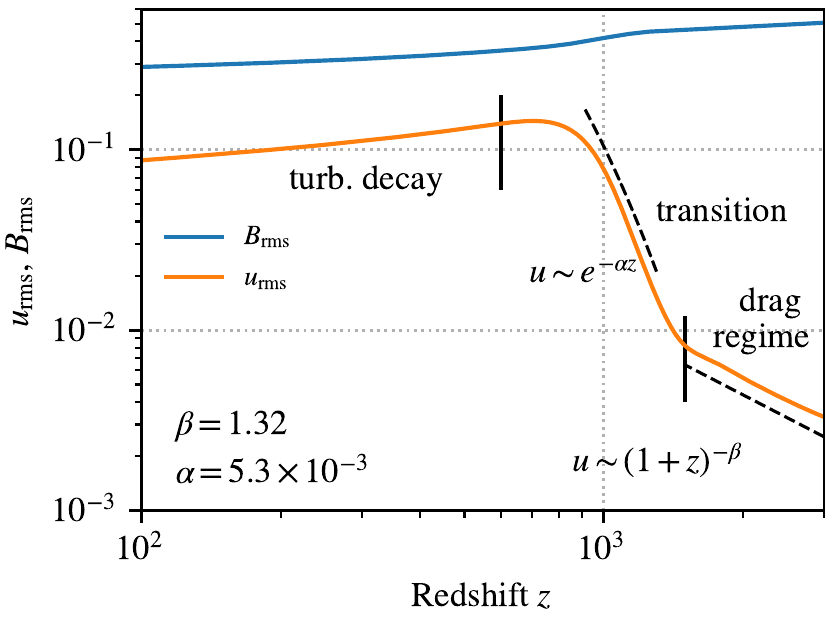}
   \caption{Evolution of the r.m.s. amplitudes (in code units, i.e. relative to the sound speed) of the magnetic field $B_\mathrm{rms}$ and baryon velocity field $u_\mathrm{rms}$ for the fiducial case of a near scale-invariant magnetic power spectrum with amplitude $B_0 = 0.51$ at $z=3000$ (corresponding to $B_{0, \rm phys} \approx 2.0 \times 10^{-2}$ nG).}
   \label{fig:tser_drag_k09}
\end{figure}

To analyze the evolution of the magnetic and kinetic energy density over the course of our simulation, we plot the root mean square (rms) velocity and magnetic field strength in Fig.~\ref{fig:tser_drag_k09}. 
At higher redshifts prior to decoupling, the fluid is dominated by photon drag.
During this drag dominated regime, there is little evolution of the magnetic field -- it decreases by less than $10\%$ from its early value of $B_0=0.51$ at $z=3000$.
The kinetic field, initially starting from zero, is still highly suppressed, by more than two orders of magnitude compared to the sound speed $c_s$, at $z=3000$. 
It then rises slowly from $u_\mathrm{rms}\approx \pot{1.5}{-3}$ to $\pot{5}{-3}$ at $z\approx 1500$ as the drag coefficient decreases (see Fig.~\ref{fig:effective_rate_k09} where the drag is shown separately). 
We fit this slower rise of the velocity field with a power law $u \, \sim \, (1+z)^{-\beta}$, with $\beta = 1.32$.

Below a redshift of $z \, \sim \, 1400$, the transition regime sets in due to a sharp decrease in the photon drag (Fig.~\ref{fig:tser_drag_k09}).
For this period we fitted the steep increase in the velocity field amplitude with an exponential function $u\, \sim \, e^{-\alpha z}$ yielding a value of $\alpha = \pot{5.4}{-3}$. 
This transition regime lasts until the velocity to magnetic field ratio approaches a maximum value of $u_\mathrm{rms}/B_\mathrm{rms} \approx 1/2$. However, actual equipartition, which one could expect, is not reached and could be related to the fields' intermittent nature over the total volume \citep{Subramanian1998turbulentdynamo,Federrath2011}. This is also similar to the departure from full equipartition observed in earlier simulations of turbulence with the \texttt{Pencil-Code} \citep[e.g.,][]{Reppin17}.

At $z\approx 800$ turbulent decay starts to set in, leading to the continued dissipation of the magnetic field energy as well as a turnover and then gradual decrease in the kinetic energy of the baryon velocity field. During the turbulent period, the r.m.s. magnetic field strength further decreases such that by $z=100$ it is approximately $60\%$ of its original value at $z=3000$. This can be seen more clearly in the upper panel of Fig.~\ref{fig:ts_comp_k09}. The transition phase captured in our simulations introduces a significant delay in the onset of turbulence.

\subsubsection{Magnetic and kinetic power spectra}

\begin{figure}
    \includegraphics{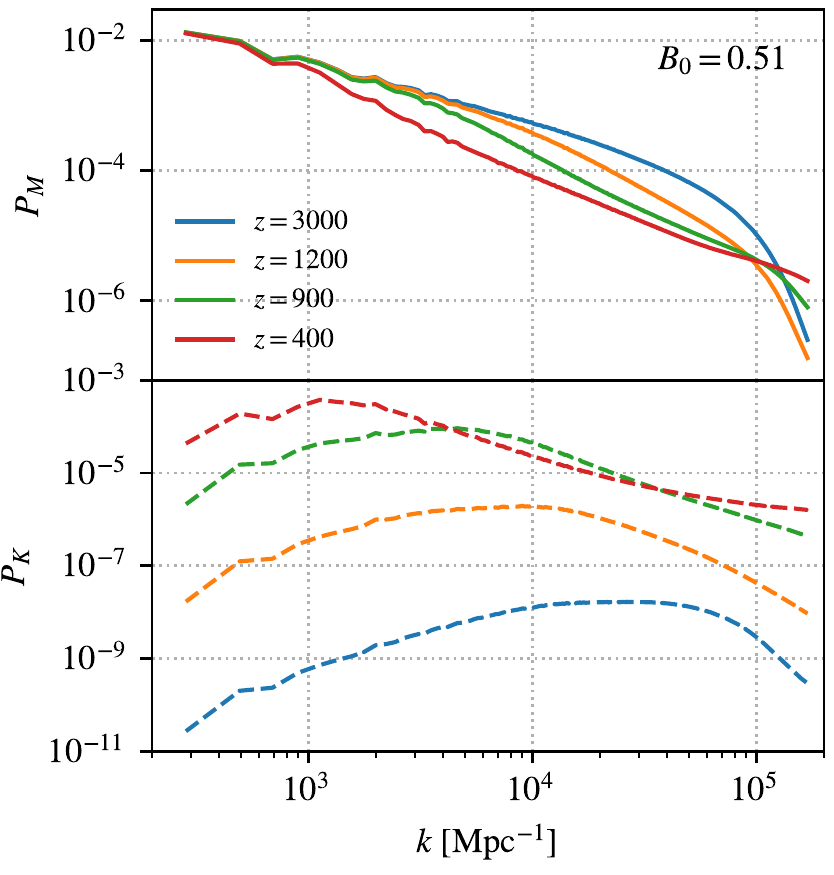}
    \caption{Evolution of magnetic (solid lines) and kinetic (dashed lines) power spectra versus redshift for a scale-invariant initial magnetic field. Spectra are shown for early times $z=3000$, during the transition regime at $z=1200\ \&\ 900$, and once the fluid has become fully turbulent at $z=400$. The magnetic field strength is $B_0 = 0.51\, \Leftrightarrow \, B_{0,\mathrm{phys}} = \pot{2.0}{-2}\mathrm{nG}$ at the reference redshift $z=3000$}
    \label{fig:mag_spectra_k09}
\end{figure}

\begin{figure*}
    \includegraphics[width=1.9\columnwidth]{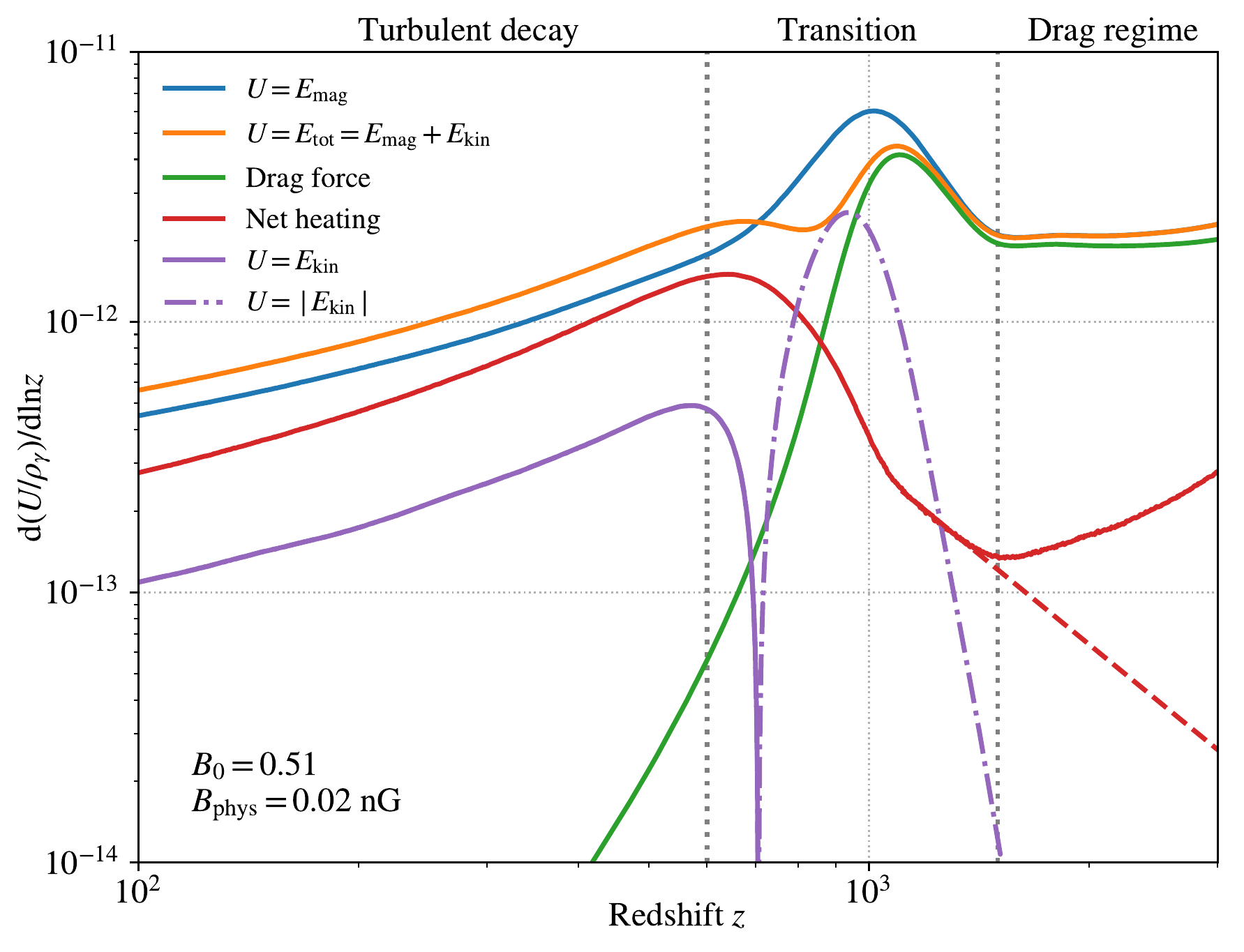}
    \caption{Energy density dissipation rates and the net heating rate (red curve) for the simulation fiducial case with a near scale-invariant spectrum $E_k\simeq k^{-0.9}$, normalized to the radiation energy density.  The net heating rate is computed via $\left[ \mathrm{d} E_{\rm tot}/\mathrm{dln} z - 2(\tilde{\alpha}+1/2) E_{\rm kin} \right]/\rho_\gamma$ (Eq.~\ref{eq:net_heating_rate_def}). The dashed red line shows the extrapolation of the effective heating rate for $\nu_3\rightarrow 0$, derived from Fig.~\ref{fig:heat_rate_highz}. The green curve shows the rate of change of the drag force $2\tilde{\alpha} E_{\rm kin}$ (without the kinematic 1/2 term). The blue and purple curves show the rate of change of magnetic and kinetic energy density using Eq.~\eqref{eq:Q_{B comp}}. The dash-dot purple line shows the negative part of the kinetic energy density dissipation rate, when the kinetic velocity field is gaining energy from the magnetic field. As the net heating rate excludes the contribution from the kinematic drag term, it does not add up to the total energy dissipation rate.}
    \label{fig:heat_rate_k09}
\end{figure*}

In Fig.~\ref{fig:mag_spectra_k09} (upper panel), we show the magnetic power spectra $P_M$ for different redshifts in our simulation for an initial near scale-invariant magnetic field ($n=-0.9$ or $n_{B}=-2.9$). 
The power spectrum $P_M$ is defined to be $\varv_A^2$ integrated over the angular variables in configuration space. 
The total energy of the magnetic field can be computed as the integral of the power spectrum in $k$-space, $E = \int P_M \dk = \int k P_M \dlnk$ and one can also define an integral scale $k_I$. 
For a scale-invariant spectrum we have $P_M\propto k^{-0.9}$ and the integral scale in our simulation is the box size $L$. 
For the more blue cases, $P_M \propto k^{n},\ n=0,1,2$ the integral scale $k_I$ is at larger $k$, a little closer to the Nyquist wavenumber in the simulation.

For the magnetic power spectra, shown in Fig.~\ref{fig:mag_spectra_k09}, the first spectrum is plotted at $z=3000$, when the ring-in phase for velocities has already ended.
The spectrum follows a power-law $P_M \propto k^{-0.9}$ up to the high-$k$ regime. 
The small-scale fluctuations have been damped out and the spectrum has a steep decline towards the Nyquist wavenumber $k_\mathrm{Ny}=N/2$, where $N$ is the resolution. 
Close to the Nyquist scale, the magnetic power spectrum has an exponential cutoff with the shape $P_M\propto  \expf{-2k^2/k_{\rm d}^2}$, with the dissipation scale $k_{\rm d}$ \citep{Jedamzik1998, Subramanian2016}

We observe that the power spectrum is slowly reshaped during the transition period, as visible for $z$=1200 and 900. This is because the ionization fraction decreases and the drag force on the baryons becomes negligible. 
The scale-invariant initial spectrum is transformed into a broken power-law spectrum with a knee below which it is steeper. 
In the turbulent phase ($z\lesssim800$), this knee, where the $k^{-0.9}$ spectrum turns into a turbulent one, moves to smaller $k$ over time, which means ever larger scales become turbulent in the simulation. 
Also, due to the negligible drag force at $z\lesssim 800$, the viscosity becomes dominated by the small hyperviscosity we added to the simulation setup. 
This has the effect of extending the spectrum more and more towards the Nyquist scale $k_\mathrm{Ny}$ and reducing the strong exponential cutoff which is apparent in the earliest spectrum shown at $z=3000$. 
At the latest redshift shown, $z=400$, the fluid has become fully turbulent at the scales present in our simulation with the power spectrum steepened to an almost Kolmogorov spectrum (with an energy distributed as $E\propto k^{-5/3}$), though in our simulations we find a power-law exponent of $\simeq -1.44$, slightly flatter than the canonical value. 

The kinetic power spectrum evolves in a clearly different manner (see the lower panel of Fig.~\ref{fig:mag_spectra_k09}). 
We start our simulation with zero initial velocity $\mathbf{u}_0=0$ and the first power spectrum of the kinetic field at $z=3000$ is during an epoch, when the fluid is heavily dominated by photon drag. 
At this epoch only a small fraction of the total energy is in the velocity field, so the power spectrum is approximately two orders of magnitude below the magnetic power spectrum. 
The kinetic power spectrum is peaked at a very large wavenumber, $k_\mathrm{peak, phys}\approx \pot{4}{4} \ (h/0.7)\ \mathrm{Mpc}^{-1}$. 

During the transition period a stronger kinetic power spectrum develops. It has a roughly $E_k\propto k^2$ shape until it reaches its peak, which shifts to lower $k$ as the kinetic power spectrum builds up. 
At wave numbers larger than the peak, there is a turbulent spectrum, with $E_k\propto k^{-5/3}$. 
By $z=900$, the slope of the turbulent part of the power spectrum is found to be $-1.71$. 
At the final redshift, $z=400$, the slope becomes slightly flatter than Kolmogorov due to a small bottleneck effect at the highest $k$-values (as tuning of hyperviscosity $\nu_3$ was done for the \textit{magnetic} power spectrum). 
As the ionization fraction decreases and becomes negligible, the turbulent part of the spectrum covers a wider range of scales, starting from lower and lower $k$-values. 
Towards the end of the simulation, $z=400$, this scale is at $k\simeq 1000\ (h/0.7)\ \mathrm{Mpc}^{-1}$. 
A video of the temporal evolution of both the magnetic and kinetic spectra can be viewed at \url{https://www.hs.uni-hamburg.de/research/cosmo-mf}.

\subsubsection{Energy dissipation rates and net heating}
In Fig.~\ref{fig:heat_rate_k09}, we show the dissipation rate, i.e the negative rate of change, per logarithmic redshift, of the magnetic energy density ${\rm d}E_\mathrm{mag}/{\rm d}\ln z$ (blue curve) calculated using Eq.~\eqref{eq:Q_{B comp}}. 
Also shown are the dissipation rates of the kinetic energy density (purple), total ($E_\mathrm{mag}+E_\mathrm{kin}$) energy density 
(orange), the photon drag force (per unit area) $2\tilde{\alpha} E_{\rm kin}$ (green), and finally the difference between the total energy dissipation rate and the drag rate, which can be interpreted as the \textit{net heating rate} being imparted into the fluid (red curve). 
Note that the kinematic drag term, related to the Hubble expansion, $-(1/2)H\mathbf{u}$, is subtracted out from the net heating and not shown as included in the drag curve (green).
At high redshifts we also show the expected net heating rate (red dashed) if we had very low viscosity at $z>1500$, taken from extrapolation from the net heating behaviour for lower $\nu_3$ shown in Fig.~\ref{fig:heat_rate_highz}.

In Fig.~\ref{fig:heat_rate_k09}, one can clearly identify the three stages discussed above: at early times, the net heating is very small and energy losses from the magnetic fields are created by the effect of photon drag. 
Starting from $z\simeq 1500$, baryon velocities build up, absorbing losses from the magnetic fields, until a turbulent flow is formed. 
Once turbulence develops below $z\sim700$, the velocity field starts losing energy via a turbulent cascade. Net heating is observed to gradually build up once baryon velocities start rising. 
The rise becomes faster as the drag decreases. 
Then the net heating peaks as the transition era gives way to turbulent decay, and net heating decreases throughout the turbulent phase down to low redshifts. 

For the chosen fiducial case initial conditions, the magnetic energy dissipation rate peaks around recombination at a redshift of about $z\approx 1000$, whereas the net heating rate peaks at $z\approx 600$. 
This significant delay ($\sim 0.5$ Myr) is caused by the time taken for the fluid to transition into the turbulent state.
The net heating rate has a broad distribution and is not a sharply peaked feature as was expected from earlier analytical studies. 
This means that the onset of modifications to the cosmological ionization history are expected to be more gradual, which will avoid numerical issues pointed out in \citet{Chluba2015PMF}. 

Note that the magnetic energy dissipation rate curve lies above the total energy dissipation rate up until a redshift of $z\approx 700$. 
This is because up until then the kinetic energy density increases, adding a negative contribution (dot-dashed purple curve) to the total energy dissipation rate. 
At late times, both magnetic and kinetic fields decay turbulently, contributing positively to the total energy dissipation rate ${\rm d}E_\mathrm{tot}/{\rm d}\ln z$. 
In this turbulent phase, the real losses from the system lead to heating of matter.

\subsection{Variation of magnetic amplitude}
\label{sec:var_B0}
In this section we present simulation results for a varying magnetic field amplitude over the range  $B_0 = (0.022-0.51)$ corresponding to $B_{\rm 0, phys} \simeq (0.086-2.0) \,\times \,10^{-2}$ nG, investigating the behaviour of magnetic and kinetic amplitudes as well as dissipation and net heating rates as a function of varying $B_0$.

\subsubsection{Evolution of magnetic and kinetic amplitude}
%
\begin{figure}
    \includegraphics{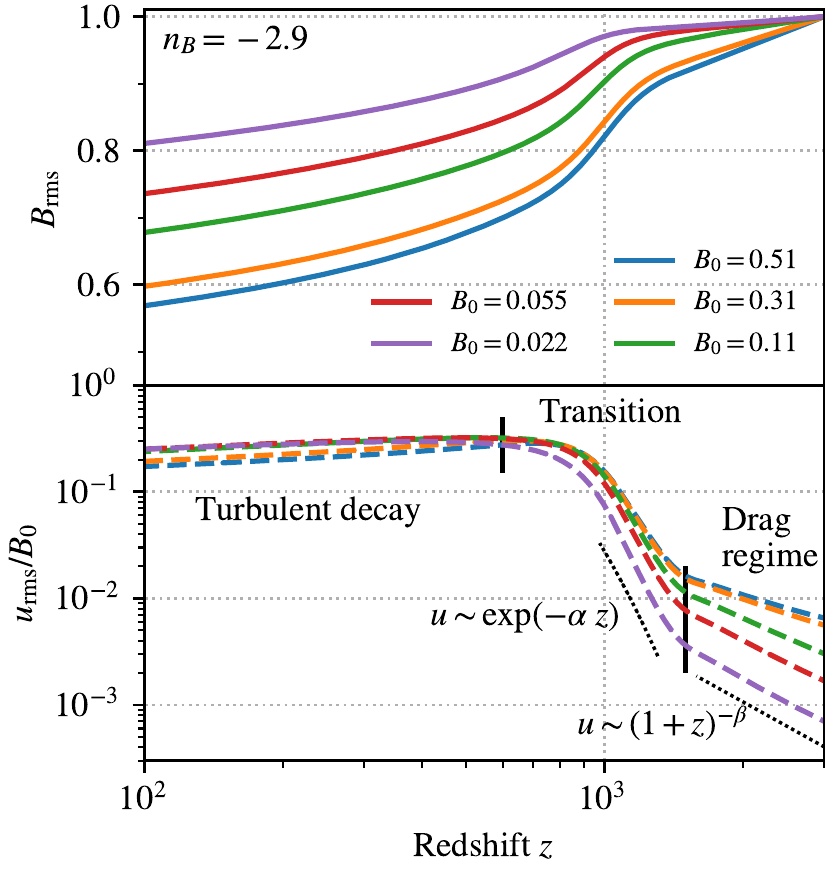}
    \caption{ Evolution of magnetic (top) and kinetic (bottom) field r.m.s. amplitudes versus redshift for different initial magnetic field strengths $B_0$. The initial field strength was varied by a factor of $\simeq 20$ (values denoted in the legend) and the magnetic amplitudes are all normalized to unity at $z=3000$. Velocities are rescaled to the magnetic field strengths $B_\mathrm{rms}$.}
    \label{fig:ts_comp_k09}
\end{figure}
%
In Fig.~\ref{fig:ts_comp_k09} we show different simulation runs with varying initial ($z=3000$) magnetic field strength amplitude $B_0$.
The upper panel shows the evolution of the magnetic field strength $B_\mathrm{rms}$ on a linear vertical scale.
The lower panel shows the velocity field strength $u_\mathrm{rms}$, rescaled to the magnetic field strength $B_\mathrm{rms}$, on a logarithmic vertical scale. For the lowest initial field strength $B_0=0.022$ there is a $20\%$ decrease of the initial rms field value over the entire redshift range  $3000> z>100$ that goes into increasing the velocity field.
This diminution or decay of magnetic fields becomes more appreciable reaching $\simeq 45\%$ for the strongest magnetic field case $B_0=0.51$, highlighting that turbulent time-scales become shorter with increasing magnetic field strength.

As we first observed for the fiducial case, we again see three distinct phases from the simulation results for different initial amplitudes. The first one is the drag dominated regime, where $u_\mathrm{rms} \ll 1$ (i.e., $\varv \ll {\rm v_{bA}}$). 
From the upper panel of Fig.~\ref{fig:ts_comp_k09}, where $B_\mathrm{rms}(z)$ is depicted, we see that for the lower $B_0$ runs, much less than a percent of magnetic field energy is transferred to $u_\mathrm{rms}$ during the drag dominated phase $3000 >z > 1500$, whereas it is $\sim 1\%$ for the highest $B_0$ case.
The $u_\mathrm{rms}$ amplitude (plotted in units of initial $B_0$) is an increasing function of the initial amplitude $B_0$ in the drag regime. 
The stronger the initial magnetic field amplitude $B_0$, the more energy is transferred to the velocity field, albeit at a relatively small fraction. 
In the subsequent transition regime, $B_\mathrm{rms}$ values decay most rapidly while the velocities rise more sharply, both effects due to the decreasing drag. 
However, by the end of the transition phase all curves catch up to reach a similar ratio of $u_\mathrm{rms}/B_\mathrm{rms} \approx 0.3$, as seen in Fig.~\ref{fig:ts_comp_k09} around $z\approx 650$. 

\begin{figure}
\centering
\includegraphics{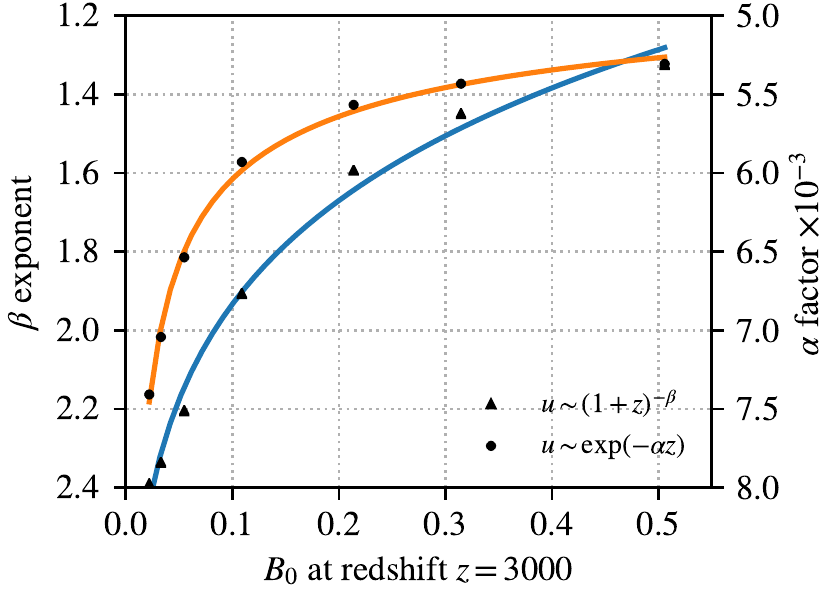}
\caption{Parameters of the fitted curves describing the evolution of the velocity field $u_\mathrm{rms}$  in Fig.~\ref{fig:ts_comp_k09}. We fitted a curve for the drag dominated regime with $u\simeq (1+z)^{-\beta}$ (triangles), one for the transition regime with $u\simeq \exp{(-\alpha z)}$ (circles). The data points were then fitted with a power-law to give an analytical representation of $\alpha$ and $\beta$, given in Eq.~\eqref{eq:alpha_beta}}
\label{fig:fit_values}
\end{figure}

We fit to the obtained velocity data (i) a power law $u \simeq (1+z)^{-\beta}$ for the drag regime and (ii) an exponential $u\simeq \exp(-\alpha z)$ for the transition regime. 
This enables us to derive simple semi-analytical fit formulae plotted in Fig.~\ref{fig:fit_values} for the parameters $\alpha$ and $\beta$ of the growing velocity field, as a function of $B_0$.
\begin{align}
 \alpha &= (4.49 + 0.57\ B_0^{-0.42})\times 10^{-3} \nonumber \\
 \beta &= 4.97 -  4.0\ B_0^{0.12} 
\label{eq:alpha_beta}
\end{align}
The coefficient $\alpha $ in the transition exponential fit is steeper for lower initial field strengths $B_0$, visible in the upper (orange) curve in Fig.~\ref{fig:fit_values}, while the drag fit power-law exponent $\beta$ plotted as the lower (blue) curve, varies less.

The last phase is that of turbulent decay where both $u_\mathrm{rms}$ and $B_\mathrm{rms}$ decrease slowly at $z<600$. This occurs for all simulation runs of the amplitude variation in a similar fashion, but with a greater overall decay for larger initial amplitude $B_0$. In the next section we compare the magnetic energy dissipation rates to those expected from turbulence. 
Although the rate of decline of $B_\mathrm{rms}$ was faster in the transition era, the overall decline in the turbulent decay regime till its end (at $z =100$ in the simulation) is comparable to the decline in the transition regime.
However, the decline in the velocity field rms amplitude over the turbulent regime, by at most a few tens of a percent up to a factor of two, is much slower than its rapid rise by more than an order of magnitude in the two earlier regimes.

%
\begin{figure}
\centering
\includegraphics{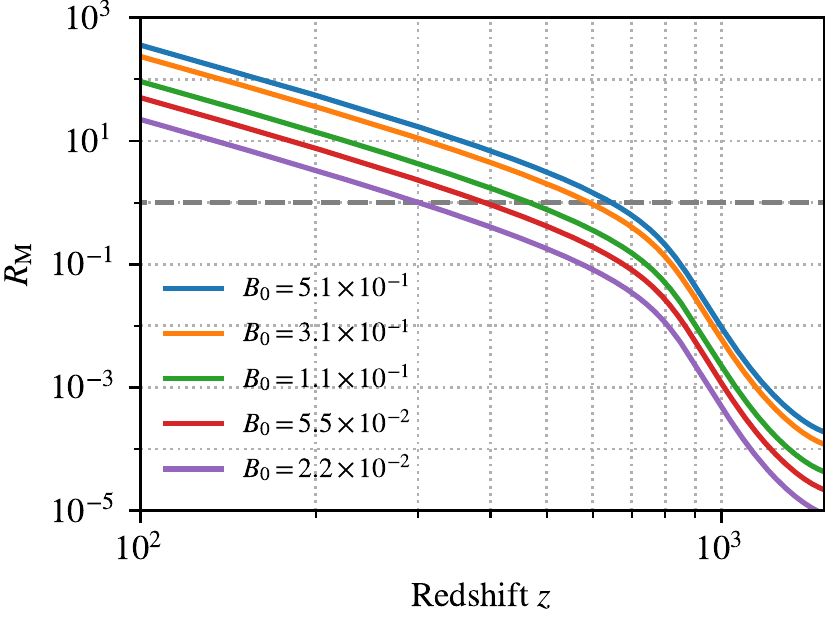}
\caption{Magnetic Reynolds numbers as a function of redshift for simulation runs where $B_0$ is varied. The critical magnetic Reynolds number $\mathrm{Rm} =1$ is indicated with a grey dashed line. For different initial amplitudes the $\mathrm{Rm} =1$ redshift varies from $z_\mathrm{td}\simeq 650$ (fiducial $B_0$ = 0.51) to $z_\mathrm{td}\simeq 300$.} 
\label{fig:reynold_amplaa_comp}
\end{figure}

The redshift at which turbulence becomes important for the velocity field is illustrated in Fig.~\ref{fig:reynold_amplaa_comp}, where we plot the evolution of the magnetic Reynolds number $\mathrm{Rm} = (\mathrm{v_A}/L)/\tilde{\alpha}$, as a function of the initial magnetic field amplitude $B_0$.
For the fiducial case, $B_0=0.51$, the magnetic Reynolds number rises above unity only at redshift $z\lesssim 650$ which matches the epoch where turbulent decay sets in for the velocity field, see Fig.~\ref{fig:ts_comp_k09}. 
For weaker fields, the transition to the turbulent regime occurs even later. 

\begin{figure}
    \includegraphics{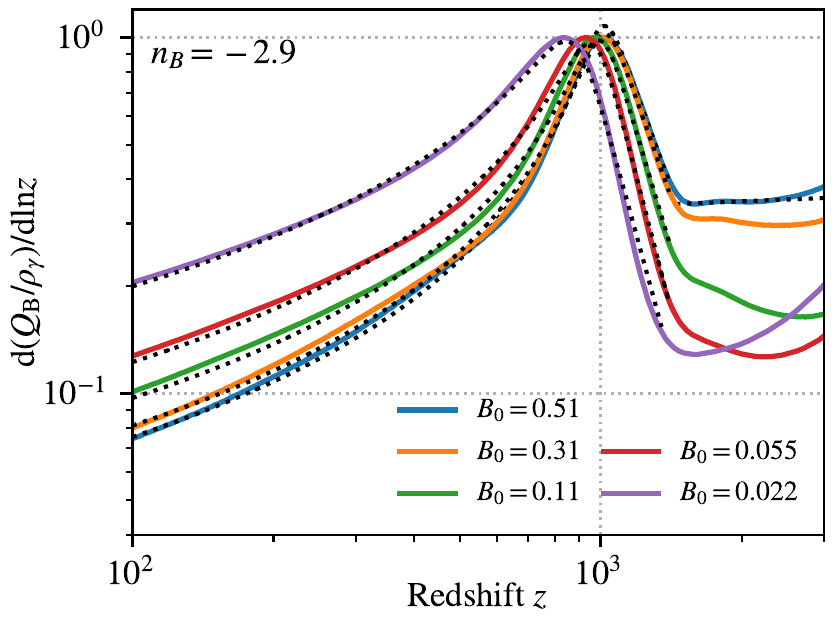}
    \caption{Comparison of the dissipation rate of the magnetic energy density with varying amplitude of the initial magnetic field $B_0$. All curves were normalized to their peak amplitude. The $z\gtrsim 1500$ drag-dominated range is affected by numerical convergence issues for the lower amplitudes. Black dotted curves are fits (Eqs.~\ref{eq:fit_func_k09_turb} $\&$ \ref{eq:fit_func_k09_trans}) for turbulent and transition regimes.}
    \label{fig:heat_rate_comp_k09}
\end{figure}

\begin{figure}
    \includegraphics{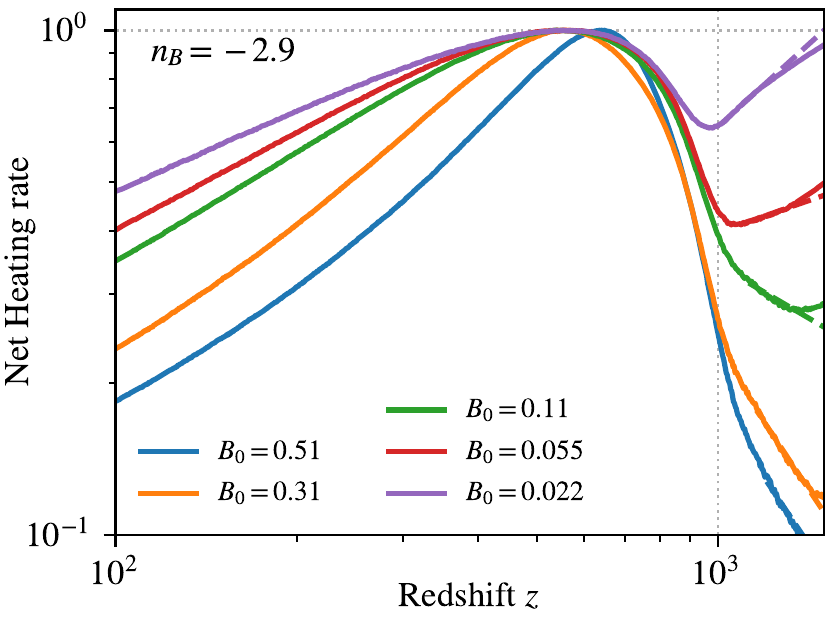}
    \caption{Net heating rates for different initial magnetic field strengths $B_0$. 
    For the high $B_0$ runs the peak of the net heating is well-pronounced. 
    The low amplitude runs are dominated by hyperviscosity at high-$z$, so dashed lines were plotted to show the physical decline of net heating for earlier times.}
    \label{fig:net_heat_rate_comp_k09}
\end{figure}
%
\begin{figure}
    \centering
    \includegraphics{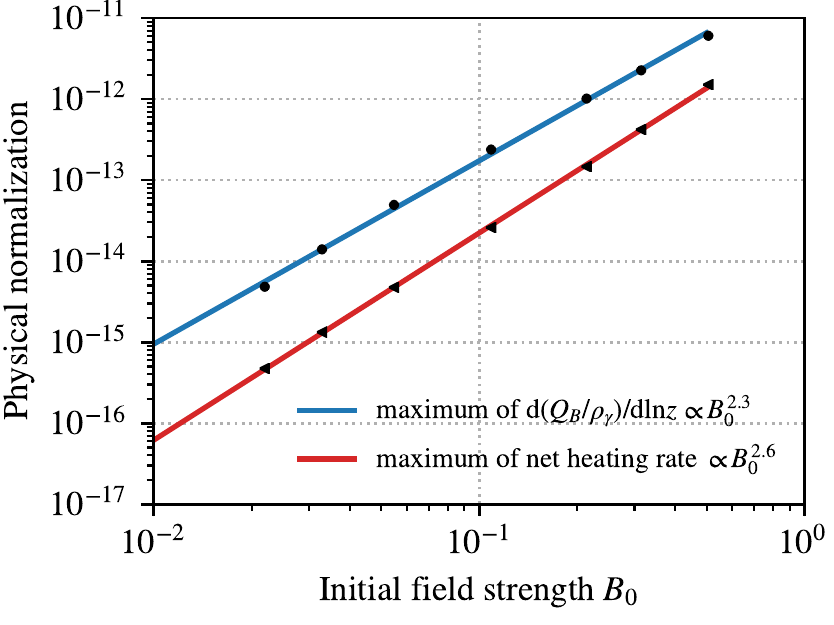}
    \caption{Physical normalization of the magnetic dissipation  rate ${\rm d}Q_B/{\rm d}\ln z$ (Fig.~\ref{fig:heat_rate_comp_k09}) at its peak (blue line) and the net heating rate (Fig.~\ref{fig:net_heat_rate_comp_k09}) at its peak (red line), for different initial field strengths $B_0$.}
    \label{fig:heat_rate_amplaa_k09}
\end{figure}

\subsubsection{Dissipation rates with varying magnetic amplitude}

We now investigate the difference in the magnetic dissipation and baryon heating rates as a function of magnetic amplitude with the aim of possibly placing constraints on primordial magnetic fields. 

The magnetic energy density dissipation rate (per logarithmic redshift interval), normalized to the energy density of radiation ${\rm d}(Q_B/\rho_\gamma)/{\rm d}\ln z $ is plotted for varying $B_0$ in Fig.~\ref{fig:heat_rate_comp_k09}. 
Note that in this plot the curves have all been normalized to their peak value, $\left[{\rm d}(Q_B/\rho_\gamma)/{\rm d}\ln z\right]_\mathrm{peak}$.
The normalization in physical units as a function of $B_0$ is given separately in Fig.~\ref{fig:heat_rate_amplaa_k09}. 
We find that there is a dependence on $B_0$ of the redshift at which the magnetic energy density dissipation peaks, particularly for the weaker fields. 
The peak redshift varies from $z\simeq 1020-820$ over the range of magnetic field amplitudes $B_0=0.51-0.022$ while exhibiting a non-linear $z_\mathrm{peak} \leftrightarrow B_0$ relation.

Another trend that was observed by varying the field strength amplitude was that the height of the peak compared to the magnetic energy dissipation rate values at $z\approx 1500$ in Fig.~\ref{fig:heat_rate_amplaa_k09} is a monotonic function of $B_0$. We find an increase of ${\rm d}Q_B/{\rm d}\ln z$ by a factor of $\simeq 3$ for $B_0=0.51$ (blue line) and an increase by a factor of $\simeq 8$ for the run with lowest amplitude with $B_0=\pot{2.2}{-2}$ (red line). The shape of $\mathrm{d}Q_B/\mathrm{d}\ln z$ between $z=3000$ and $z=1500$ is less robust and also more susceptible to numerical issues so we ignore those values, since we do not extract any physical properties from the simulation during this period. For the $B_0=0.51$ the high-$z$ behaviour can be described by a power-law
\begin{equation}
 \left.\frac{\mathrm{d}(Q_B/\rho_\gamma)}{\mathrm{d}\ln z}\right|_\mathrm{drag}= 0.23\times (1+z)^{0.054}.
\end{equation}
For the lower amplitudes the numerical noise is too strong at those early redshift, so we do not infer anything from the high-$z$ behaviour.
For the magnetic dissipation rate, the peak position does not coincide with Rm $\simeq 1$, but rather with the epoch at or just before when magnetic and kinetic energy densities reach closest to equipartition (see Fig.~\ref{fig:ts_comp_k09}). 
The real importance of turbulence can only be appreciated when considering the net heating rates (see Fig.~\ref{fig:net_heat_rate_comp_k09} and discussion below), which show that for a large redshift range
the PMF evolution for small magnetic fields is only weakly turbulent. 
This leads to a net heating rate with a very broad maximum, in contrast to previous analyses.

\begin{figure}
\centering
\includegraphics{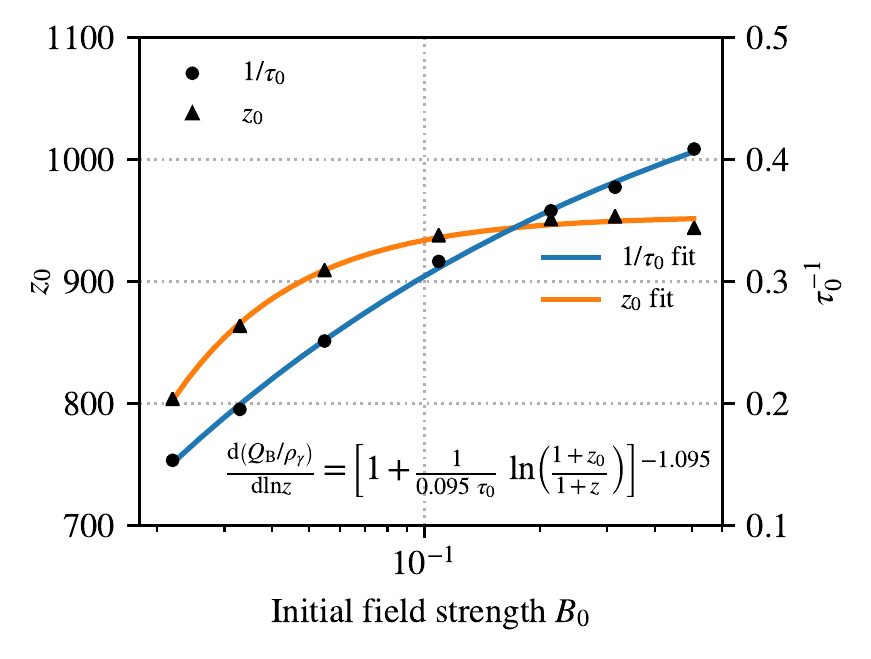}
\caption{Fitting parameters for the semi-analytic formula used to describe the magnetic dissipation rates (Fig.~\ref{fig:heat_rate_comp_k09}) in the turbulent decay regime for different $B_0$. The redshift parameter $z_0$ saturates towards $z \sim 950$ whereas the parameter $1/\tau_0$ increases for larger $B_0$.}
\label{fig:amplaa_fitting_paras}
\end{figure}

The net heating rate is computed by subtracting the dissipation due to drag, which can be calculated as $2(\tilde{\alpha}+1/2) E_{\rm kin}$, from the total energy dissipation rate ${\rm d}E_\mathrm{tot}/{\rm d}\ln z$. 
In Fig.~\ref{fig:heat_rate_k09} we had displayed the rates of $E_{\rm mag},\ E_K,\ E_\mathrm{tot},$ drag dissipation rate and the net heating rate for our fiducial case, $\nu_3=\pot{2.5}{-16},\ B_0=0.51$. 
We  now show the net heating rates $\left[ \mathrm{d} E_{\rm tot}/\mathrm{dln} z - 2(\tilde{\alpha}+1/2) E_{\rm kin} \right]/\rho_\gamma$ for the simulation runs with varying magnetic amplitudes in Fig.~\ref{fig:net_heat_rate_comp_k09}.
The net heating rate peaks at similar redshifts $z\sim 550-650$, with most of the small shift in epoch occurring between $B_0$=0.51 and 0.31. It rises to a peak over a roughly similar redshift range towards the end of the transition regime, once the drag dissipation has become negligibly small, and the peak epoch does not strongly dependent on varying $B_0$. The peak epoch of the net heating is significantly later ($\sim 0.5$ Myr) compared to the peak of magnetic dissipation, for all $B_0$ values, as already mentioned above, due to the time taken to develop sufficiently turbulent velocity fields.
Another new result from these simulations is that net heating generally does not peak sharply but over a broad range of redshifts, increasingly so for weaker initial amplitudes $B_0$.
The strongest amplitude case $B_0=0.51$ (blue) lowest curve in Fig.~\ref{fig:net_heat_rate_comp_k09}) has the most pronounced peak at $z\approx 650$ in the heating. For the lower $B_0$ runs the peak is very broad and almost constant over a considerable range of redshifts $z\sim 500-700$.

In Fig.~\ref{fig:heat_rate_amplaa_k09} we show the physical normalizations of the amplitude of the peak of ${\rm d}Q_B/{\rm d}\ln z$ (blue line), as well as the maximum value of the net heating rate (red line) as a function of the initial magnetic field strength $B_0$.
We find that the magnetic energy dissipation rate has the following dependence of the initial magnetic field in our simulations:
\bsub 
\begin{align}
  \left. \frac{\mathrm{d} (Q_B/\rho_\gamma)}{\mathrm{d} \ln z }\right|_\mathrm{peak} = 3.17\times 10^{-11}\times B_0^{2.26}\\
\left.\frac{\mathrm{d}Q_\mathrm{net \,\,heat}}{\mathrm{d}\ln z}\right|_\mathrm{peak} = 8.04\times 10^{-12}\times B_0^{2.55}.
\end{align}
\esub
Since the magnetic energy dissipation rate is simply the logarithmic derivative of the magnetic energy density $\propto B^2$, we expect a scaling of ${\rm d}Q_B/{\rm d}\ln z\propto B_0^2$.
The observed scaling of the amplitude in our runs is slightly steeper than quadratic.

\subsubsection{Semi-analytic representation of magnetic dissipation}
We also show fits to the turbulent and transition regime portions of the magnetic dissipation rates as a function of $B_0$. For the low redshift turbulent decay dominated part, we fit an inverse logarithmic function of the form derived phenomenologically in Eq.~\eqref{eq:QbarB_IV}, for the case of a near scale-invariant magnetic spectrum $n_B=-2.9 \Leftrightarrow m = 0.095$:
\begin{equation}
 \left. \frac{\mathrm{d} (Q_B/\rho_\gamma)}{\mathrm{d} \ln z }\right|_{\mathrm{turb}} = \left(1 + \frac{1}{(0.095)\tau_0} \ln\frac{1+z_0}{1 +z}\right)^{-1.095}.
 \label{eq:fit_func_k09_turb}
\end{equation}
This fitting process for parameters $\tau_0$ and $z_0$ was applied over a range of low redshifts $z \in [100, z_\mathrm{max}]$, where $z_\mathrm{max}$ ranges from $500$ to $900$, for initial amplitude $B_0 = 0.51-0.022$. 
This fitting range was chosen to give better agreement in the lower redshift part of the turbulent regime. 
The resulting parameters for the semi-analytic magnetic energy dissipation rate are shown in Fig.~\ref{fig:amplaa_fitting_paras}. 
By applying a power-law fit to the obtained values we can give a description of the parameters $z_0$ and $\tau_0^{-1}$ as a function of the initial magnetic field strength $B_0$:
\begin{align}
 z_0 &= 959 -2.37\, B_0^{-1.08} \nonumber \\
 \tau_0^{-1} &= 0.75 -0.30\, B_0^{-0.17}.
\end{align}
We should note that the fitted coefficients and exponents given above are somewhat sensitive to the chosen redshift range in which the fit is performed. 

In the transition regime, on the high redshift side of the maximum rate of change of the magnetic energy density, we fitted a power-law of the form:
\begin{equation}
 \left. \frac{\mathrm{d}(Q_B/\rho_\gamma)}{\mathrm{d}\ln z}\right|_\mathrm{trans} = \,\,A\times (1+z)^{-\beta},
\label{eq:fit_func_k09_trans}
\end{equation}
where $\beta \in [3.6, 5.0]$. The two separate fits over the turbulent and transition regimes were co-added with a cubic exponent and over-plotted as black dotted lines in Fig.~\ref{fig:heat_rate_comp_k09},
\begin{equation}
 \left. \frac{\mathrm{d}(Q_B/\rho_\gamma)}{\mathrm{d}\ln z}\right|_\mathrm{drag} = \,\,A\times (1+z)^{\gamma}.
\label{eq:fit_func_k09_drag}
\end{equation}
With $A=0.23$ and $\gamma = 0.054$.
For $B_0=0.51$. The simulation  of the lower amplitudes $B_0$ have not fully equilibrated at high-$z$ and the behaviour of $\mathrm{d}Q_B/\mathrm{d}\ln z$ is dominated by numerical effects, resulting in slopes $\gamma$ that show non-monotonic behaviour as a function of $B_0$, so we only fit for the run with highest $B_0$. 

\subsection{Dependence on spectral index}
\label{sec:var_nB}
Since we set up the magnetic field power spectrum as an initial condition, we have the freedom to vary its spectral shape. A scale-invariant magnetic power spectrum $P_M \propto k^{-0.9}$ is expected from models of inflationary magnetogenesis \citep{Subramanian2016}.
We also know that $P_M \propto k^4$ is an initial power spectrum that represents a magnetic field generated in a causal process, like a first-order phase transition \citep{Durrer03}. Probing PMF evolution for different magnetic spectra thus has the potential to constrain and discern among different models of magnetogenesis.
In the following subsections we describe the trend of physical results from simulation runs with varying spectral indices. 

\subsubsection{Evolution of magnetic and kinetic field amplitudes}
%
\begin{figure}
    \includegraphics{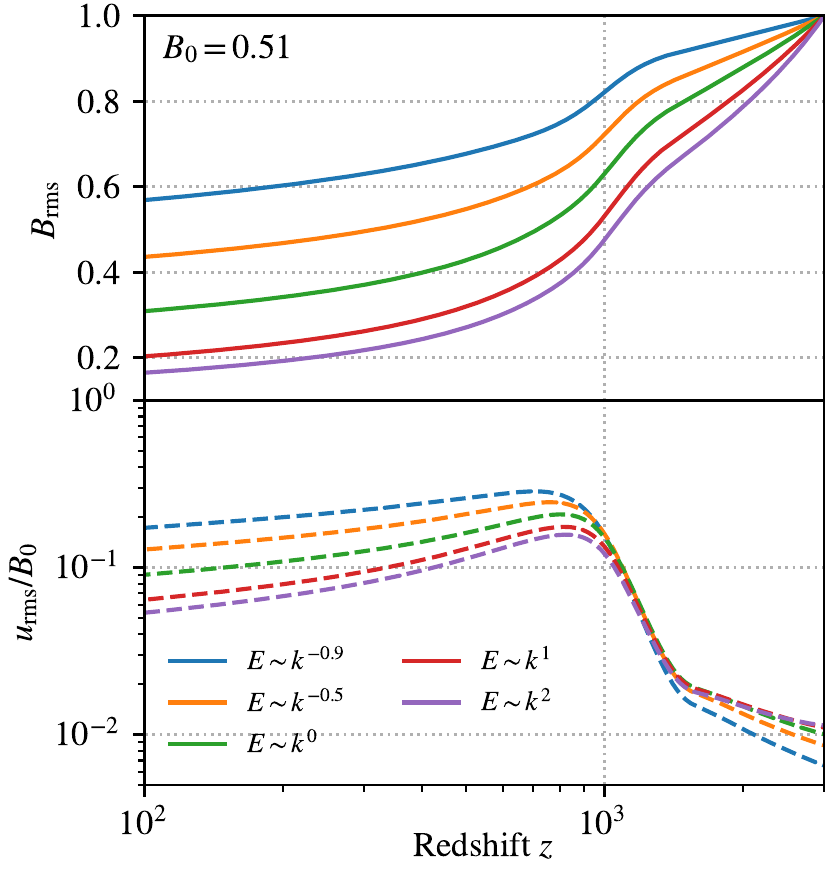}
    \caption{Evolution of the rms amplitudes with different initial spectra. The solid lines give the magnetic field strength, the dashed lines are the \textit{rms} velocity in the simulation. For the steep spectra the same initial integral scale was used.}
    \label{fig:ts_comp_index}
\end{figure}
%
In Fig.~\ref{fig:ts_comp_index} we show the redshift evolution of the magnetic field $B_\mathrm{rms}$ in the upper panel and the velocity field $u_\mathrm{rms}$ in the lower panel. 
The different curves correspond to (from top to bottom in the upper panel) a near scale-invariant spectrum $P_M\propto k^{-0.9}$ (blue), two flatter spectra $P_M\propto k^{-0.5}$ (orange), $P_M\propto k^0$ (green) and two steep spectra with $P_M\propto k^1,\ P_M\propto k^2 $ which are plotted as red and purple lines, respectively. 
The curves for $B_\mathrm{rms}$ and $u_\mathrm{rms}$ were normalized at $z=3000$, when the ring-in phase of the velocity field died down.
It should be kept in mind that the relative evolution between $z=3000$ and the starting redshift of the simulation depends on the spectral index.

For the magnetic field strength $B_\mathrm{rms}$ there is a clearer difference in its evolution as a function of its spectrum.
The scale-invariant case shows the least decay: its $B_\mathrm{rms}$ value at $z=100$ is $60\%$ of the initial value at $z=3000$, with a greater fraction of the decay happening during the later phases $z<1000$.
On the other hand, the steepest spectrum $P_M\propto k^4$ decays by almost an order of magnitude during the simulation, with more than half the decline occurring prior to $z = 1000$.
In general, there seems to be a clear distinction between the $B_{\rm rms}$ evolutionary shape for the scale-invariant case and the steeply evolving most blue spectra.
Steepening the spectral index among the blue spectra ($k^1, k^2$) only slightly changes their magnetic amplitude evolution and they exhibit a very similar decay shape.

For the velocity field there is no strong dependence on the spectral index of the initial magnetic field at high-$z$ in the drag and transition regimes. 
All curves show a similar slope and amplitude except the scale-invariant fiducial case where velocities grow slightly faster.
Only at lower redshifts, below $z<800$ and in the turbulent regime do the differences become apparent.
The larger the initial spectral index $n$ is, the closer one gets to equipartition, defined in our units as equality of $u_\mathrm{rms}$ and $B_\mathrm{rms}$.
Also, the runs with steeper initial spectra show more pronounced turbulent decay of velocities.

\subsubsection{Magnetic power spectra}

\begin{figure}
    \includegraphics{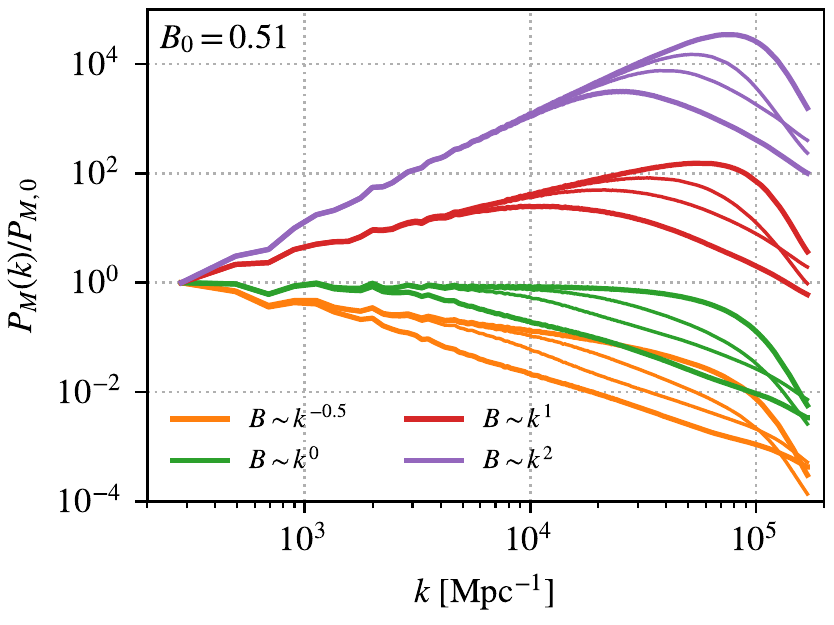}
    \\[1mm]
    \includegraphics{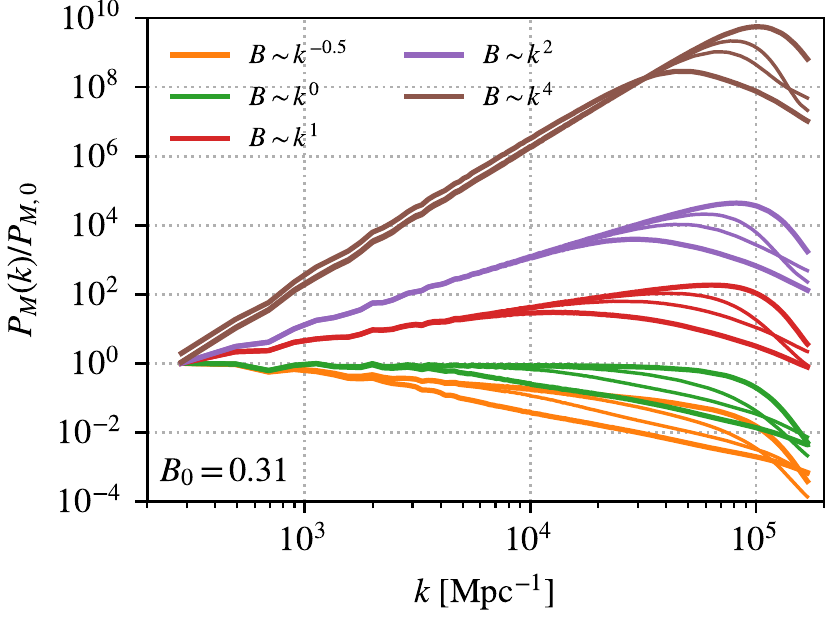}
    \caption{Evolution of the magnetic power spectra with different spectral index $n$, at four redshifts $z = 3000, 1200, 900\ \&\ 400$, decreasing from top to bottom in each panel. The upper panel is for $B_{z=3000} = 0.51$ while the lower shows the curves for $B_{z=3000} = 0.31$. For $B_{z=3000}=0.31$, a causal initial spectrum, $P_M\propto k^4$, is also simulated and shown.}
    \label{fig:spectra_comp_index}
\end{figure}

\begin{figure}
    \includegraphics{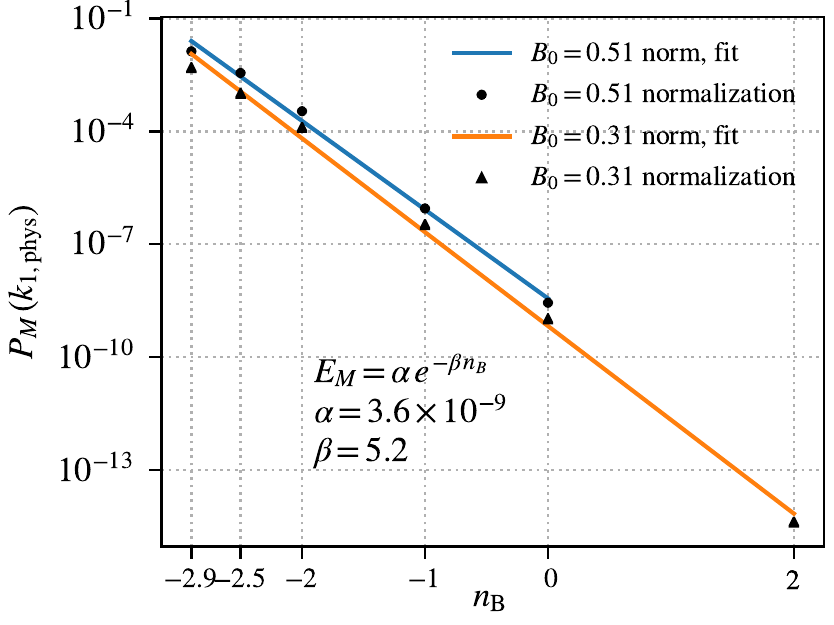}
    \caption{Relative normalization of the magnetic power spectra (shown in Fig.~\ref{fig:spectra_comp_index}) at $k_{1,_\mathrm{phys}} = 280\,(h/0.7)$ Mpc$^{-1}$ as a function of spectral index $n_B = n-2$}.
    \label{fig:spectra_index_norm}
\end{figure}
We show the magnetic power spectra $P_M(k)/P_M(0)$ in Fig.~\ref{fig:spectra_comp_index} evolving through the same redshifts $z=3000,1200,900\ \&\ 400$ as in Fig.~\ref{fig:mag_spectra_k09} but now for a range of initial spectral indices. 
These spectra can be contrasted with the evolving spectra already presented for our fiducial case $B_0=0.51$, $n=-0.9$ in Fig~\ref{fig:mag_spectra_k09}.
We observe that for $B_0=0.51$ (upper panel of Fig.~\ref{fig:spectra_comp_index}), the spectrum with $P_M\propto k^{-0.5}$ 
shows an evolution of the spectrum into a turbulent power-law over virtually all its $k$-modes. However, this is not the case for steeper spectral indices.
For the $P_M\propto k^{0}$ case, the lowest wavenumber at which the initial spectrum has evolved into a turbulent Kolmogorov-like spectrum $P_M\simeq k^{-5/3}$ is $k\approx \pot{3.0}{3} \, (h/0.7)\ \mathrm{Mpc}^{-1}$. 
The power spectrum case of $P_M\propto k^1$, the reddest spectrum for which an integral scale $k_I$ can be defined, displays
$k_I\simeq  \pot{3.8}{4} \, (h/0.7)\ \mathrm{Mpc}^{-1}$ at $z=400$. 
For the $P_M\propto k^2$ the integral scale at $z=400$ shifts to wavenumber $k_I\approx \pot{4.8}{4} \,(h/0.7)\ \mathrm{Mpc}^{-1}$.
All the integral and peak scales for bluer spectra shown in Fig.~\ref{fig:spectra_comp_index} are presented in Table~\ref{tab:spectra_peaks_integralscales}.

In Fig.~\ref{fig:spectra_comp_index} (lower panel) 
we show the magnetic power spectra for the runs with a lower initial magnetic field strength of $B_0=0.31$, where the steepest case of an initial causally-generated power spectrum, $P_M\propto k^4$, is also presented. 
We observe that this case too develops a turbulent high-$k$ tail for $k>k_I$ with $k_I\approx \pot{6.4}{4} \, (h/0.7)\ \mathrm{Mpc}^{-1}$.
The evolution of power spectra presented in Fig.~\ref{fig:spectra_comp_index} clearly demonstrate the establishment of a turbulent power-law over time, whose $k$-range is largest for scale-invariant spectra $k^{-0.9}$ and progressively restricted only to higher $k$-scales as the spectra become more blue to $k^2$ and $k^4$.
The blue spectrum simulation runs also exhibit the start of an inverse transfer effect for decaying non-helical turbulence.
\cite{Reppin17} have shown that this can occur for steep spectra if Reynolds numbers are large enough and the integral scale $k_I$ is peaked at sufficiently large $k$, which is the case here \citep[also cf.][]{Brandenburg2015}.

The normalization of magnetic power spectra has been set by relating its integral to the chosen $B_0$.
The relative normalization of the magnetic power spectra as a function of spectral index $n_B=n-2$ is shown in Fig.~\ref{fig:spectra_index_norm}. 
The normalization is computed at the first binned wave vector value of $k_1=1.29$ which corresponds to $k_{1,\mathrm{phys}}= 280\,(h/0.7)$ Mpc$^{-1}$. 
The dependence of this normalization on the spectral index $n_B$ follows an exponential form and Fig.~\ref{fig:spectra_index_norm} also gives a fit.

\begin{table}
 \centering
\begin{tabular}{ccc}
 \toprule
Magnetic spectrum & \,\,\,integral scale $k_I$ \,\,\,& peak scale\\
 \midrule
 $\!\!\!\!\!\!\!\!\!\!\!\!\!\!\!\!\!\!\!\!\!\!B_0 =0.51$ & &\\
 $n=1$ & $\pot{3.8}{4}$ & $\pot{9.0}{3}$  \\ 
 $n=2$ & $\pot{4.8}{4}$ & $\pot{2.4}{4}$  \\ 
 \midrule
$\!\!\!\!\!\!\!\!\!\!\!\!\!\!\!\!\!\!\!\!\!\!B_0 =0.31$ & &\\
$n=1$ & $\pot{4.0}{4}$ & $\pot{1.2}{4}$  \\ 
 $n=2$ & $\pot{5.1}{4}$ & $\pot{2.8}{4}$  \\ 
 $n=4$ & $\pot{6.4}{4}$ & $\pot{4.6}{4}$  \\ 
 \bottomrule
\end{tabular}
\caption{The integral and peak scale in physical units of $(h/0.7)\ \mathrm{Mpc}^{-1}$, for the more blue magnetic spectra, measured at $z=400$ (spectra shown in Fig.~\ref{fig:spectra_comp_index})}
\label{tab:spectra_peaks_integralscales}
\end{table}

\begin{figure}
    \includegraphics{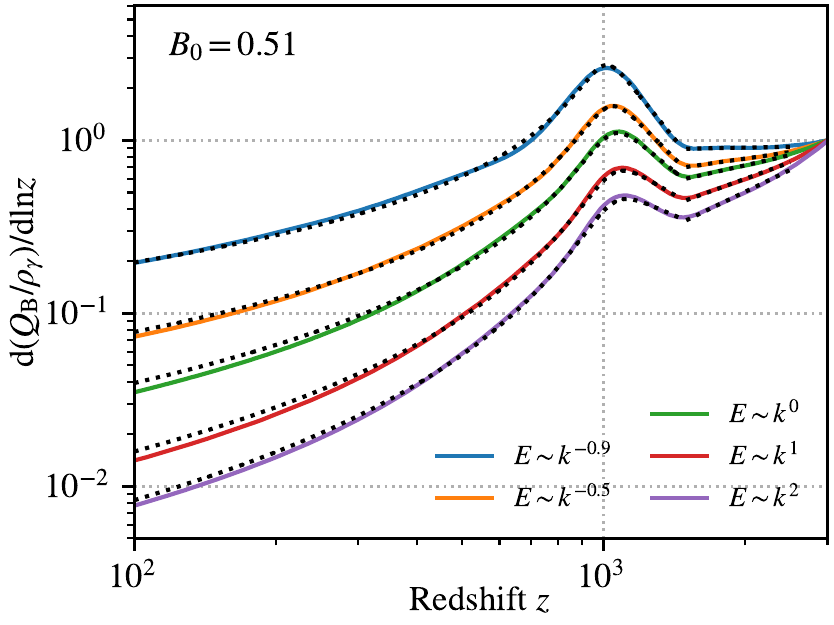}
    \caption{Comparison of dissipation rate of magnetic energy density when varying the spectral index $P_M k\propto k^n$, with $n=n_B+2$. The curves are normalized to the values at $z=3000$. The spectral index varies from $-0.9$ to $4$ from top curve to bottom. We show combined fits (dotted curves) to the magnetic energy dissipation rates for the drag regime at high-$z$ (Eq.~\ref{eq:drag_param}), for the turbulent regime (Eq.~\ref{eq:QbarB_temp_sols}) and for the transition region (Eq.~\ref{eq:QbarB_transition_temp}).}
    \label{fig:heat_rate_comp_index}
\end{figure}

\begin{figure}
\centering
\includegraphics{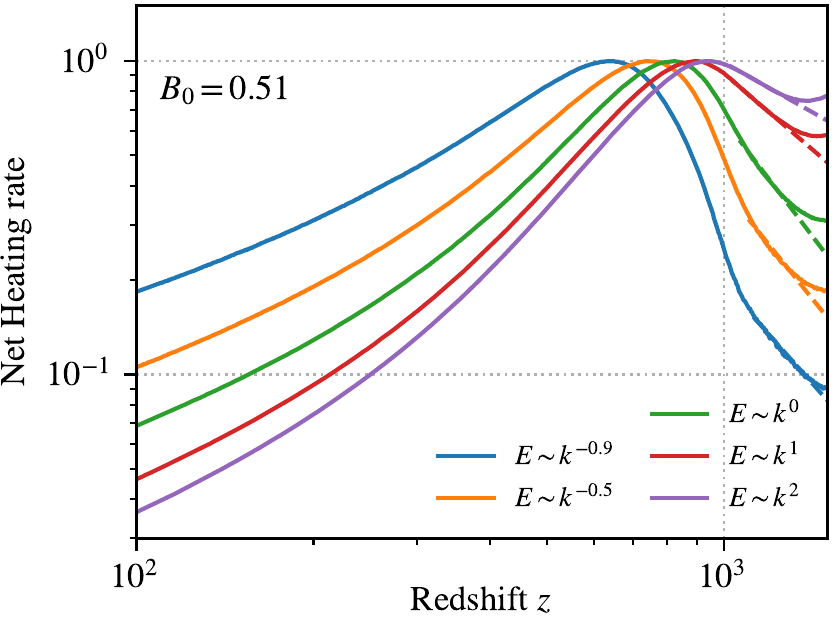}
\\[1mm]
\includegraphics{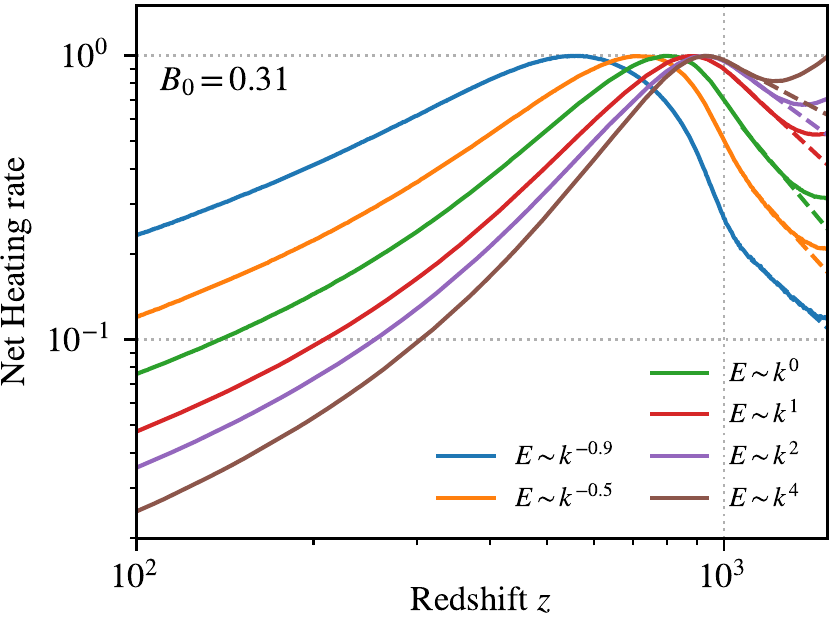}
\caption{Net heating rate for different spectral indices $n$ and two values of $B_0$ =0.51 and 0.31. The physical normalizations for both cases are presented in Fig.~\ref{fig:index_scalings}. The dashed lines show an extrapolation of the rates avoiding the numerical issues which dominate at large $z$ and cause an unphysical rise, especially of the bluer spectra.}
\label{fig:net_heat_index_comp}
\end{figure}

\begin{figure}
\centering
\includegraphics{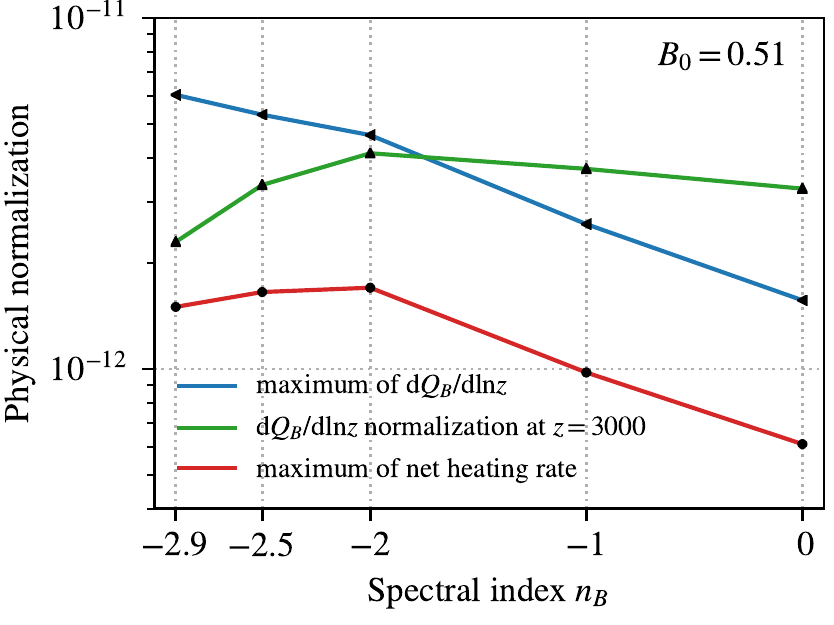}
\caption{Physical normalization of the peak values (blue) and initial (z$=3000$) values (green) of the magnetic energy density dissipation rate from Fig.~\ref{fig:heat_rate_comp_index} as well as the net heating rate peak values (red) from Fig.~\ref{fig:net_heat_index_comp}, all as a function of spectral index $n_B$ where $n_B=n-2$.}
\label{fig:index_scalings}
\end{figure}

\subsubsection{Dissipation rates and net heating rate}
\label{sec:total_PMF_loss_n}
We now examine the changes in the magnetic energy dissipation rate and the net heating rate for the simulation runs with varying spectral indices. 
Fig.~\ref{fig:heat_rate_comp_index} depicts the rate of change of the magnetic energy density. Note that here we have normalized all curves to their initial value at $z=3000$. 
The peak redshift is virtually independent of spectral index $n$, tending slightly to earlier epochs for bluer spectra. 
In the drag dominated regime $z > 1500$, the slope of the energy dissipation rate becomes monotonically steeper with increasing $n$. 
This has the effect that the magnetic dissipation rate only rises above its initial value for spectra redder than $P_M \propto k^0$. In the turbulent regime, the dissipation rates all decline monotonically, by over a factor of 3 for the near scale-invariant spectrum and by around an order of magnitude for the bluest spectrum. We fit to each of these three regimes as described in the next subsection.

The net heating rate due to magnetic fields for different initial spectral indices $n$ = -0.9, -0.5, 0, 1 $\, \& \,$ 2 is displayed in Fig.~\ref{fig:net_heat_index_comp}. Two different magnetic field amplitudes $B_0$=0.51 and 0.31 (including the additional spectral case of $n=4$) are shown in the upper and lower panels respectively.
The epoch of peak net heating shows a striking shift to higher redshifts as the spectral index varies from $n$=-0.9 to 2 or 4. 
There is a clear distinction between the scale-invariant case $n=-0.9$, which peaks at a redshift $z\approx 650$, and the net heating for steeper initial spectra, which peak at earlier epochs up to $z \approx 925$. 
Thus magnetic spectral variation alone can delay the epoch of peak net heating by $\sim$ 0.4 Myr. 
This effect is even more  pronounced as $B_0$ is reduced from 0.51 to 0.31, as expected from the amplitude variation of net heating described earlier in Fig.~\ref{fig:net_heat_rate_comp_k09}.

Overall, we discern two clear trends for the shape and position of the net heating rate as a function of magnetic field parameters: from earlier, as the field strength grows stronger, the net heating maximum is more sharply peaked in time. 
The second trend we see here is that as the magnetic spectral index is lowered (to near scale-invariant), the net heating peak is delayed by a greater time interval of up to 0.5 Myr after the epoch of recombination. To compare, this time-lag is greater than the age of the Universe at recombination.

\begin{figure}
 \centering
 \includegraphics{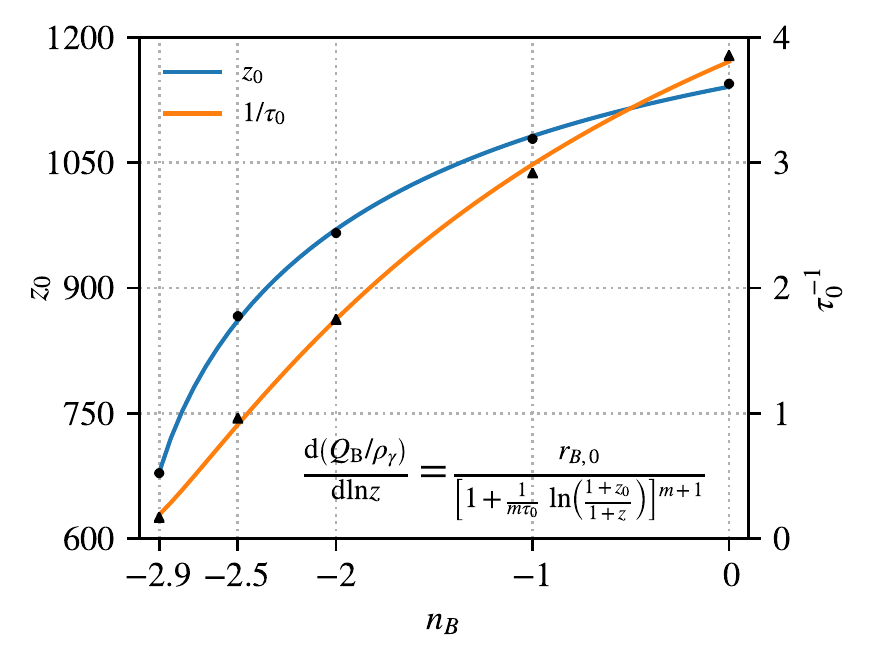}
 \caption{Fitting parameters for the magnetic energy density dissipation rate with varying spectral index in Fig.~\ref{fig:heat_rate_comp_index} in the turbulent decay regime. The rates were fitted using Eq.~\eqref{eq:QbarB_temp} over the range $z\in[100,900]$. A power-law was then fitted to give a functional form of $z_0$ and $\tau_0^{-1}$ in Eq.~\eqref{eq:QbarB_temp_sols}.}
\label{fig:dQBdlnz_comp}
\end{figure}

\subsubsection{Semi-analytic representation of the PMF losses}
\label{sec:semi-analytics}
In this section, we construct a semi-analytic representation of the total magnetic energy density dissipation rates. We provide separate fits for the turbulent, transition and drag-dominated regimes which we then combine together to plot as the black dotted curves in Fig~\ref{fig:heat_rate_comp_index}.

In the turbulent regime, following the phenomenological derivation presented in Sect.~\eqref{sec:turb_decay}, we use the functional form
\begin{align}
\label{eq:QbarB_temp}
\left.\frac{\id (Q_{B}/\rho_\gamma)}{\id \ln z}\right|_{\rm turb}\approx \frac{r_{B, 0} }{\left[1+\frac{1}{m\tau_0}\ln\left(\frac{1+z_0}{1+z}\right)\right]^{m+1}},
\end{align}
with free parameters $\tau_0$ and $z_0$ as a starting point. We recall that $m=2(n_B+3)/(n_B+5)=2(n+1)/(n+3)$. The late-time behavior of the curves shown in Fig.~\ref{fig:heat_rate_comp_index} can then be represented by 
\bsub
\label{eq:QbarB_temp_sols}
\begin{align}
r_{B, 0}&\approx 1
\\
z_0&=532 + 549\,m^{0.56}
\\
\tau^{-1}_0&= 0.07 + 2.91\,m^{1.36}.
\end{align}
\esub
Since the denominator in Eq.~\eqref{eq:QbarB_temp} vanishes at $z_{\rm lim}\simeq z_0\,\expf{\,m\, \tau_0}$, Eq.~\eqref{eq:QbarB_temp} is only valid at $z<z_{\rm lim}$, where $z_\mathrm{lim} \in [1190\ ; 1560]$, and the turbulent regime lies well within this range. The power-law fits for $z_0$ and $\tau_0^{-1}$ are plotted in Fig~\ref{fig:dQBdlnz_comp} as a function of spectral index $n_B$

Next, the transition regime between drag-dominated ($z\gtrsim 1500$) and pure turbulent-decay ($z\lesssim 800$), can be represented using a simple power-law scaling:
\begin{align}
\label{eq:QbarB_transition_temp}
\left.\frac{\id (Q_{B}/\rho_\gamma)}{\id \ln z}\right|_{\rm trans}\approx A\,\left[\frac{1300}{1+z}\right]^\beta.
\end{align}
To represent the rising dissipation rates from the drag regime to their peak values in Fig.~\ref{fig:heat_rate_comp_index}, we find
\bsub
\begin{align}
\label{eq:QbarB_transition_temp_sols}
A&\approx 1.59 - 1.04\,m^{0.64}\\
\beta&=3.51 - 1.94 \, m
\end{align}
\esub
to work very well. 
For the transition regime the power-law was fitted in a redshift range of $z \in [1500,1200]$.

\begin{figure}
 \centering
 \includegraphics{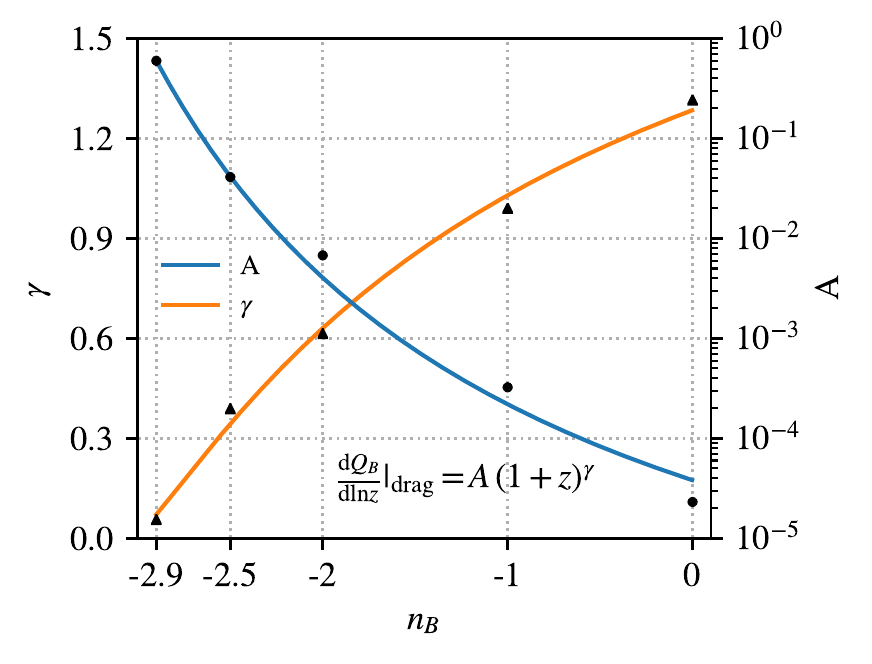}
 \caption{Fitting parameters for the high-$z$ drag-dominated magnetic energy dissipation rate in Fig.~\ref{fig:heat_rate_comp_index}, as a function of spectral index. The fit equations for the power-law parameters $A$ and $\gamma$ are given in Eq.~\eqref{eq:drag_param} }
 \label{fig:high_z_fits}
\end{figure}

To approximate the evolution at the high redshift drag-dominated regime, we fit a power-law of the following form
\begin{equation}
 \left.\frac{\mathrm{d}(Q_B/\rho_\gamma)}{\mathrm{d}\ln z}\right|_\mathrm{drag} = A \left(1+z\right)^\gamma
\end{equation}
to the magnetic energy dissipation rate in Fig.~\ref{fig:heat_rate_comp_index} for $z\in [1500,\, 2500]$. 
This gives the observed slopes shown in Table~\ref{tab:drag_slopes}.
If we parameterize the slope $\beta$ and the amplitude $A$ as a function of the initial spectral index $m$
we get:
\bsub
\label{eq:drag_param}
\begin{align}
 A &= 1.38\, e^{-8.75\, m} \\
 \gamma &= 0.015 + 1.01\, m^{1.23}.
\end{align}
\esub 
We normalize the curves at $z=3000$. Fig.~\ref{fig:high_z_fits} shows the parameterization of the fits to the high redshift range in Fig.~\ref{fig:heat_rate_comp_index}~.

In the pre-recombination drag dominated regime, the theoretical prediction derived in Sect.~\ref{sec:drag_analytic} for the magnetic energy density dissipation rate $\id Q_{B}/\mathrm{dln}\, z \propto \rho_\gamma(z) \,(1+z)^{5(n_B+3)/2(n_B+5)}$ is qualitatively comparable to the trend of the fitted slopes $\gamma$ as a function of spectral index - see Table~\ref{tab:drag_slopes}~. 
We note that our numerical simulations do not explicitly match the predicted high-$z$ scaling (cf. Eq.~\ref{eq:B2_fs}) because of numerical effects (discussed in Appendix~\ref{sec:appendix1}) and also due to our assumption of matter-domination (see discussion in Sect.~\ref{sec:pic_pre_rec}).

For a combined fit for the dissipation rates for each spectral index over all three regimes, we found that a cubic exponent gives a good approximation for combining the turbulent and transition fits at lower redshifts,
\begin{align}
\label{eq:QbarB_tot}
\frac{\id (Q_{B}/\rho_\gamma)}{\id \ln z}\approx
\begin{cases}
\left[\left.\frac{\id (Q_{B}/\rho_\gamma)}{\id \ln z}\right|_{\rm turb}^{-3}+\left.\frac{\id (Q_{B}/\rho_\gamma)}{\id \ln z}\right|_{\rm trans}^{-3}\right]^{-1/3}& z\lesssim z_{\rm lim}
\\[2mm]
\left.\frac{\id (Q_{B}/\rho_\gamma)}{\id \ln z}\right|_{\rm trans}& z\lesssim 1300
\\[2mm]
{\rm max}\left(\left.\frac{\id (Q_{B}/\rho_\gamma)}{\id \ln z}\right|_{\rm trans},
\left.\frac{\id (Q_{B}/\rho_\gamma)}{\id \ln z}\right|_{\rm drag}\right)& z\gtrsim 1300.
\end{cases}
\end{align}
With this we obtain the fit functions over plotted as the dotted black curves shown in Fig.~\ref{fig:heat_rate_comp_index}.

\begin{table}
 \centering
\begin{tabular}{cccc}
 \toprule
 $n_B$ & observed slope & fitted $\gamma$ & predicted slope\\
 \midrule
 -2.9 & 0.05 & 0.07 & 0.12 \\ 
 -2.5 & 0.39 & 0.34 & 0.50 \\
 -2.0 & 0.61 & 0.63 & 0.83 \\
 -1.0 & 0.99 & 1.03 & 1.25 \\
 0.0 & 1.31 & 1.28  & 1.50 \\
 \bottomrule
\end{tabular}
\caption{Comparison of the observed (Fig.~\ref{fig:heat_rate_comp_index}) and fitted slope $\gamma$ (Eq.~\ref{eq:drag_param}) versus the predicted slope $5(n_B+3)/2(n_B+5)$ for the dissipation rate of magnetic energy density in the high redshift drag-dominated regime.}
\label{tab:drag_slopes}
\end{table}

\section{Discussion}
\label{sec:discussion}
We discuss below some of the caveats of our computations. There are related to i) resolution and numerical viscosity, ii) baryon density perturbations and shocks for $B_0 > 10^{-2}\,{\rm nG}$, iii) self-gravity and photon perturbations, iv) changes induced to the drag coefficient. 

Let us start with resolution issues. Our technique of solving the MHD equations on a grid means that we are limited to about three orders of magnitude in dynamic range in $k$-space with our resolution of $N=1536^3$. However, we can still adequately trace the evolution of the power spectra as well as the individual kinetic and magnetic energy densities. Going to higher resolution will greatly increase the computational costs but it will also increase the range in $k$-space that can be resolved. This would allow us to separate the drag-dominated damping scale more cleanly from the Nyquist scale, hence reducing numerical artifacts. 

At the highest redshifts ($z>3000$) of our simulation, the evolution of the energy dissipation rates is still affected by numerical viscosity leading to a ring-in phase of the velocity field and a slight underestimation of the magnetic dissipation rate (since part of the energy branches into numerical heating). However, we performed several tests in this regime, strongly reducing the numerical viscosity. This showed that the PMF evolution is well understood and by the transition epoch, numerical viscosity no longer significantly affects our heating rates. Thus, we are able to reliably deduce the net heating rates that go into the plasma from the decay of the magnetic energy density at $z\simeq 3000$ to $z\simeq 100$.

In our simulation setup we are restricted to sub-sonic Alfv\'en velocities. This limits us to $B_\mathrm{0,phys}\lesssim \pot{4}{-2} \mathrm{nG}$ for physical values of the magnetic field in our runs. While still physically interesting and instructive, this level of PMFs is more than one order of magnitude below the level currently constrained using CMB data, $B_\mathrm{0,phys} \sim 1\,\mathrm{nG}$ \citep{Planck2015PMF}.
In this regime of stronger magnetic fields, one has to consider significant perturbations in the plasma as well as shocks, which we have ignored in our case of low Mach numbers. We expect this to increase the heating efficiency of the PMFs, with kinetic energy being converted into heat more rapidly. We could expect that the transient regime becomes shorter and the net heating might also evolve accordingly. This will have further implications for structure formation, so a simulation that takes such effects into account will be needed for the future treatment of heating due to stronger magnetic fields.

Stronger magnetic fields could also source significant additional baryon density perturbations which can grow after recombination. While we have ignored these in the present simulation involving relatively weak magnetic fields, they could play a role with stronger magnetic fields: altering the Jeans scale for structure formation \citep{Subramanian1998,Sethi2005} as well as potentially constraining the PMF via contribution to the small-scale matter power spectrum and further effects \citep{Shaw:2010ea,Jedamzik2013,Pandey2015,Jedamzik2018}.

Similarly, we neglected self-gravity and photon perturbations in our simulation. In the drag-dominated regime, small-scale photon perturbations will be created. Part of these perturbations could affect the subsequent evolution of the plasma in the turbulent phase. They will also become visible as ultra-small scale CMB anisotropies, well below the photon diffusion scale $k\gtrsim 0.2\,\Mpc^{-1}$. At late stages, we also expect the growth of dark matter halos to influence the dynamics of the fluid flow. This will be particularly important at $z\lesssim 100-200$, when ambipolar diffusion should also become relevant \citep{Jedamzik1998}. In refined treatments of the problem, these effects should be considered carefully. 

In our simulations, we used the pre-computed standard recombination history obtained with {\tt CosmoRec} \citep{Chluba2010b} to determine the drag coefficient. However, significant heating by PMFs can directly affect the recombination history \citep[e.g.,][]{Sethi2005, Chluba2015PMF}, which in turn would modify the drag terms. We expect these effects to only become noticeable for higher PMF strength ($B_0\simeq1 \,{\rm nG}$), when shocks also have to be considered. Therefore, the heating rates obtained here should be relatively robust with respect to this aspect.

Finally, in our simulations we only considered non-helical magnetic fields. However, PMFs may have been generated with substantial helicity \citep{Vachaspati2001} which would significantly alter their subsequent evolution \citep{Banerjee2004} as well as produce distinct cosmological signatures \citep{Pogosian2002,Planck2015PMF,Ballardini2015,Long2015}. The influence of magnetic helicity conservation as well as inverse cascade on the evolution and decay of PMFs in this case would be interesting to explore and is left for future work.

\section{Conclusions}
\label{sec:conclusion}
We performed 3D numerical simulations of PMFs along with baryon velocity fluctuations across the cosmological recombination era. Photon drag was included using the standard recombination history obtained with {\tt CosmoRec} and the MHD equations were solved in an expanding medium. Our simulations allow us to trace the flow of energy from the magnetic field, through the baryon velocity field to heating via turbulent decay and dissipation. We are able to describe the net heating rates smoothly across the epoch of recombination. This has enabled a clean separation of real heating, which will lead to an increase of the matter temperature, from drag-dominated magnetic energy losses to the CMB photon field, which only lead to secondary CMB temperature perturbations and spectral distortions.

We supported our computations with analytic estimates and provide several useful expressions to represent our numerical results. We find that at redshifts below $z\lesssim 500$, in the regime of well-developed turbulent decay, our new analytic approximation eq.~\eqref{eq:QbarB_IV} with an additional $1/m$ factor gives a good fit.

Three main evolutionary stages for magnetic heating are observed (Fig.~\ref{fig:heat_rate_k09}): i) an initial phase that is dominated by photon drag ($z\gtrsim 1500$), ii) an intermediate transition period around cosmological recombination, when the photon drag force drops rapidly, and iii) the final fully turbulent MHD phase ($z\lesssim 600$). Only in the later part of the transition and the turbulent phase do we find significant heating of the medium, a result that is important when deriving and interpreting constraints on PMFs from CMB measurements.

Our computations show that the growth of baryon velocity fluctuations is  strongly suppressed at early times because of high photon drag, but builds up during the intermediate transition phase (Fig.~\ref{fig:ts_comp_k09}). After recombination, a velocity saturation state approaching equipartition between magnetic and kinetic energy is reached, after which the plasma dissipates energy in a turbulent cascade. 
In previous treatments, the build-up of the velocity field and its turbulent decay were assumed to proceed instantaneously or in a prescribed fashion at recombination ($z\simeq 1100$). 
Our simulations reveal a substantial delay for the onset of turbulence until redshift $z\simeq 600-900$ (this depends on the amplitude and shape of the initial magnetic power spectrum). 
In the transition phase, the magnetic field decay continuously sources the baryon velocity field until a well developed turbulent state is reached, a process that causes a delay of up to 0.5 Myr. 

We also find that the shapes of the net heating rates obtained are broader than previously estimated semi-analytically. 
For weaker magnetic fields, their peak broadens to extend to significantly lower redshifts ($z\lesssim 400$). 
In these cases, the fluid flow reaches turbulence more gradually with the magnetic Reynolds number exceeding unity only in the very late stages (Fig.~\ref{fig:reynold_amplaa_comp}). 
In contrast, steeper spectra ($n \sim 1-4$) produce an earlier peak in the net heating rates at  $z_\mathrm{peak}\simeq 900$. 
The shape, epoch and width of the peak of the net heating rate are seen to depend significantly on the initial amplitude of the magnetic fields and their spectral index (Fig.~\ref{fig:net_heat_rate_comp_k09} and Fig.~\ref{fig:net_heat_index_comp}). 
Our simulations further reveal that the peak magnetic energy density dissipation rate occurs once the drag coefficient has already dropped appreciably and the fluid approaches its maximum kinetic energy density growth rate (Fig.~\ref{fig:ts_comp_k09}).

For the near scale-invariant case, almost the entire magnetic power spectrum (over two orders of magnitude) is reshaped to a turbulent power-law (Fig.~\ref{fig:mag_spectra_k09}) as a turbulent cascade removes and redistributes power. The turbulent slope obtained is $n \sim -1.4$, slightly shallower than Kolmogorov-type turbulence. On the other hand, we find that only the intermediate-scale wave modes $k>\pot{1.0}{4}\ \mathrm{Mpc}^{-1}$ are reprocessed for spectra with $P_M\propto k^1$ and only the smallest scale modes $k> \pot{5.0}{4}\ \mathrm{Mpc}^{-1}$ for the steepest spectrum $P_M\propto k^4$ (Fig.~\ref{fig:spectra_comp_index}).
The direct observation of the spectral transformation from an initial power-law with a cut-off to a spectrum in its transitional phase and then to a processed spectrum with a turbulent slope is a novel feature of our computations and a benefit of the full 3D treatment of the MHD equations.

To be able to achieve a more complete cosmological treatment, more physics at late times would need to be included: below a redshift of $z\lesssim 200$ effects like ambipolar diffusion could dominate and can contribute substantially to the evolution of the magnetic heating history at later epochs. Stronger magnetic fields along with compressible fluid phenomena would extend the scope of these simulations. A simultaneous treatment of density perturbations induced by PMFs would complement the evolution of heating and its cosmological effects.
The results presented above represent a first step in simulating magnetic heating in cosmology and it would be interesting to extend them to describe magnetic signatures and constraints arising from the CMB, structure formation, 21-cm signals as well as the epoch of reionization.

\vspace{-0mm}
\small
\section*{Acknowledgments}
JC would like to cordially thank Ethan Vishniac and Geoff Vasil for useful discussions about magnetic turbulence. We would also like to thank Jacques Wagstaff for his contribution at an early stage of the project and Stefan Hackstein for comments.
PT and JR are supported by the German Deutsche Forschungsgemeinschaft (DFG), Sonderforschungsbereich (SFB) 676 section C9. JC is supported by the Royal Society as a Royal Society University Research Fellow at the University of Manchester, UK.


\small 
\vspace{-0mm}
\bibliographystyle{mn2e}
\bibliography{Lit}

\vspace{-0mm}
\begin{appendix}
\section{Numerical viscosity tests}
\label{sec:appendix1}

\begin{figure}
    \includegraphics{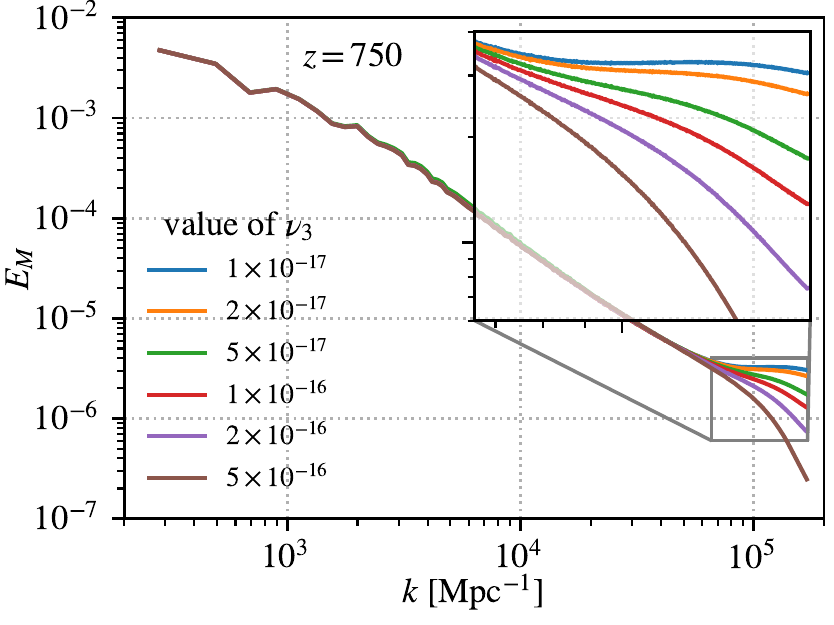}
\caption{Comparison of magnetic power spectra for varying hyperviscosity $\nu_3$. Using a value $\nu_3< 2\times 10^{-16}$ gives excess power at high $k$. This is called the bottleneck effect which we avoid by chosing $\nu_3=2.5\times 10^{-16}$}
    \label{fig:heat_rate_visc_comp_inset}
\end{figure}

\begin{figure}
    \centering
    \includegraphics{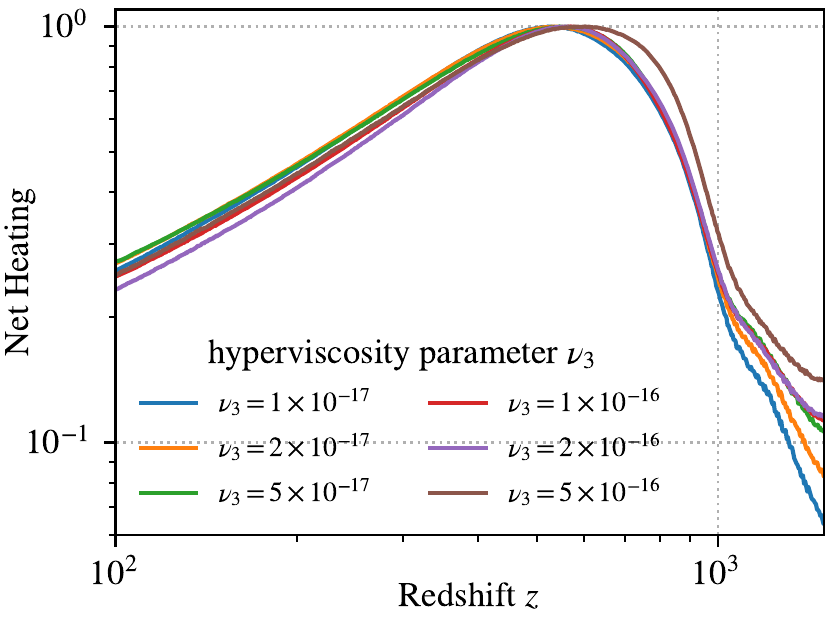}
    \caption{Dependence of the net heating rate for different $\nu_3$ runs. The heating rate goes toward zero for low viscosities. We expect this behavior since the only dissipation at high-$z$ should be from the drag. This curve is used to extrapolate the physical net heating rate for high-$z$ drag-dominated regime.}
    \label{fig:heat_rate_highz}
\end{figure}

To achieve a stable simulation over a wide range of redshifts, we tuned the hyperviscosity $\nu_3$ in such a way that the bottleneck effect does not show up at late times ($z<800$).
Due to the drag being large at early times the simulation is stabilized and the magnetic field is frozen-in, whereas at late times hyperviscosity is needed in order to have a stable simulation run.
In Fig.~\ref{fig:heat_rate_visc_comp_inset} we show the magnetic power spectra for $B_0=0.31$ at a redshift of $z=750$ for different hyperviscosity parameters $\nu_3$ ranging from $\nu_3= \pot{1}{-17}$ to $\nu_3=\pot{5}{-16}$.
We chose a value of $\nu_3=\pot{2.5}{-16}$ to get a power-law over a large range of $k$ modes.
Also, the highest value $\nu_3=\pot{5}{-16}$ already shows an exponential cutoff at $k\approx k_\mathrm{Ny}$, which causes the power spectrum not to be a smooth power-law.
We wanted to produce as little dissipation as needed, so the highest value we tested in our runs turned out to be too large.

To ensure the results we obtain in our simulations do not depend on the viscosity parameter $\nu_3$ in our 
setup, we ran multiple runs with varying runtime parameters for $\nu_3$. 
In Fig.~\ref{fig:heat_rate_highz} we plot the net heating rate for different $\nu_3$ for the amplitude $B_0=0.31$.
At high redshift $z>1000$ we see that there is some dependence of low values of the heating rate on the numerical viscosity. 
We expect the net heating rate to be very low at high $z$ since we are in the drag dominated regime and all dissipation should be due to the drag force.

\section{Resolution Study}
 
\begin{figure}
 \centering
 \includegraphics{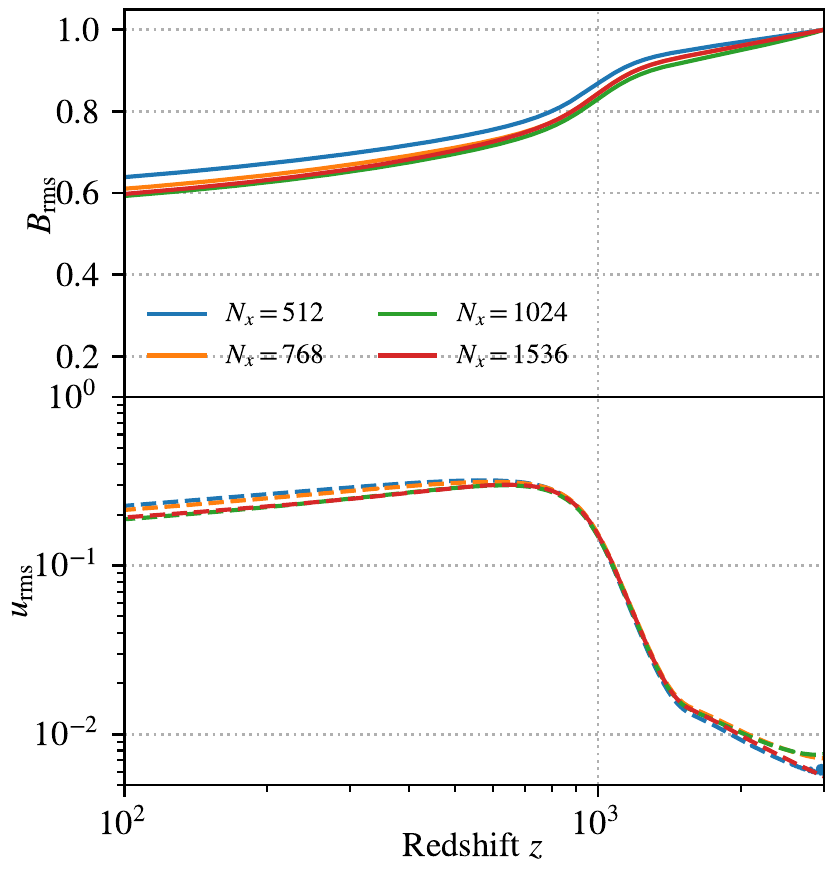}
 \caption{Redshift evolution of the magnetic field strength $B_\mathrm{rms}$ and the velocity field $u_\mathrm{rms}$ as a function of the resolution. The high resolution runs converge very well and even for the low resolution there is agreement on the qualitative behaviour.}
 \label{fig:ts_resolution}
\end{figure}

We demonstrate the convergence of our simulation with respect to different choices of numerical resolution.
For this resolution convergence test the setup with $B_0=0.31$, $\nu_3=\pot{2.5}{-16}$ and a near scale-invariant initial magnetic power spectrum with four resolutions $N=512^3,\ N=768^3,\ N=1024^3,\ \&\ N=1536^3$ were run.
In Fig.~\ref{fig:ts_resolution} we show the redshift evolution of velocity \& magnetic field strength $u_\mathrm{rms}\ \&\ B_\mathrm{rms}$ at late times, between $z=3000$ and $z=100$.
We see that the runs converge nicely, with only the two lowest resolution runs with $N=512^3$ and $N=768^3$ showing a slight deviation from our \textit{fiducial} run with $N=1536^3$.

\begin{figure}
 \centering
 \includegraphics{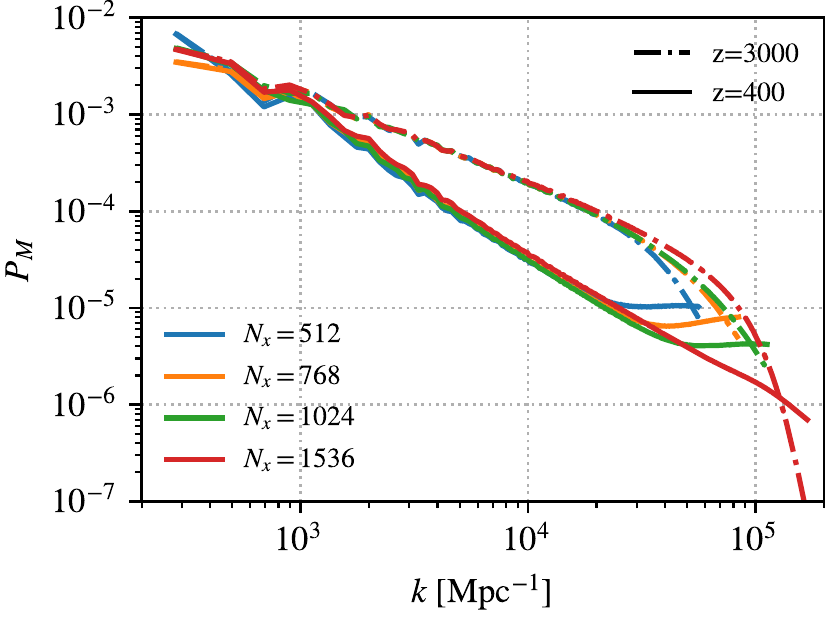}
 \caption{Magnetic power spectra at early ($z=3000$) and late ($z=400$) times for simulation runs with different resolution.}
 \label{fig:spec_resolution}
\end{figure}

In Fig.~\ref{fig:spec_resolution} we show the power spectra of these different runs at early times $z=3000$ as well as late times $z=400$. 
The initial spectra differ slightly from each other since we start the simulation at $z>3000$ and have a \textit{ring-in} phase that is not shown here.
The two runs with highest resolution converge reasonably well.
For the runs with lower resolution there is a pile-up of magnetic power at $k\approx k_\mathrm{Ny}$. 
This is the so-called bottleneck effect and shows that energy is not dissipated fast enough and the viscosity pre-factor $\nu_3$ should be higher, for that particular resolution, to increase dissipation.
For our chosen highest resolution of $N=1536^3$, this bottleneck does not occur at a significant level and we obtain a power-law spectrum over all scales.
%
%
\end{appendix}

\end{document}